\documentclass[letterpaper,journal]{IEEEtran}

\usepackage{cite}
\usepackage{amsmath,amssymb,amsfonts}
\usepackage{pifont}    
\usepackage{mathtools}
\usepackage{xcolor}
\usepackage{textcomp, gensymb} 
\usepackage{gensymb}
\usepackage{bm}
\usepackage{wasysym}
\usepackage{siunitx}
\usepackage{booktabs}
\usepackage{comment}
\usepackage[export]{adjustbox}
\usepackage[normalem]{ulem}

\newcommand\new[1]{\textcolor{blue}{#1}}

\DeclareRobustCommand{\new}[1]{\begingroup\color{blue}#1\endgroup}

\newcommand{\Internal}{0} 

\newcommand{\revisionStatus}[1]{%
  \ifnum\Internal=1
    \ifnum#1>0 {\color{blue}Draft \checkmark}\else {\color{gray}Draft}\fi, 
    \ifnum#1>1 {\color{blue}1st revision \checkmark}\else {\color{gray}1st revision}\fi, 
    \ifnum#1>2 {\color{blue}2nd revision \checkmark}\else {\color{gray}2nd revision}\fi, 
    \ifnum#1>3 {\color{blue}Complete \checkmark}\else {\color{gray}Complete}\fi
    \\
  \fi
}

\newcommand\tick[1]{\makebox[0pt][l]{$\square$}\raisebox{.10ex}{\hspace{0.1em}$\checkmark$}}

\usepackage{tikz}
\usepackage{lipsum}

\newcommand{\highlights}[2][left]{%
  \ifnum\Internal=1
  \begin{tikzpicture}[remember picture, overlay]
    \node[rotate={\ifnum\pdfstrcmp{#1}{left}=0 90\else -90\fi}, anchor={\ifnum\pdfstrcmp{#1}{left}=0 east\else west\fi}, font=\color{blue}\scriptsize, align=center] at ({\ifnum\pdfstrcmp{#1}{left}=0 -1-0.4\else 1+8.4\fi}, 0.4) {%
      \parbox{4cm}{#2}};
  \end{tikzpicture}%
\fi
} 

\usepackage{amsthm} 
\theoremstyle{definition} 
\newtheorem{theorem}{Theorem} 
\newtheorem{remark}{Remark}   
\newtheorem{assumption}{Assumption}
\newcommand{\remarkend}{\hfill\(\qed\)}%
\newcommand{\theoremend}{\hfill\(\blacksquare\)}

\usepackage[font=footnotesize,labelfont=bf]{caption}
\usepackage{subcaption}

\usepackage{textcase} 

\makeatletter
\let\NAT@parse\undefined
\makeatother
\usepackage[
    colorlinks=true,      
    linkcolor=blue,       
    urlcolor=blue,        
    citecolor=blue,       
    pdfborder={0 0 0},    
]{hyperref}

\usepackage{orcidlink}

\usepackage{titling}
\usepackage{fancyhdr}

\usepackage{titling}

\pretitle{%
  \vspace{-1.8em}%
  \noindent\rule{\textwidth}{2pt}\par\vspace{1em}
  \centering\LARGE\bfseries
}
\posttitle{%
  \par\vspace{0.2em}%
  \noindent\rule{\textwidth}{0.8pt}\par\vspace{0.5em}
}

\title{\LARGE \bf 
Cooperative Control of Hybrid FES-Exoskeleton: Dynamic Allocation
}

\date{}  
\author{Hossein~Kavianirad,~Satoshi~Endo,~Davide~Astarita,~Lorenzo~Amato,~Emilio~Trigili,~and~Sandra~Hirche}

\begin{document}

\maketitle

\thispagestyle{fancy}
\pagestyle{fancy}

\fancyhf{} 
\renewcommand{\headrulewidth}{0pt} 
\fancyhead[C]{\footnotesize This is the authors’ electronic preprint version of the article.}
\fancyfoot{} 

\begingroup
\renewcommand\thefootnote{} 
\footnotetext{This work was supported in part by the European Research Council~(ERC) Consolidator Grant ”Safe data-driven control for human-centric systems~(CO-MAN)” under grant agreement number 864686, and the Horizon 2020 research and innovation program of the European Union under grant agreement no. 871767 of the project ReHyb.}
\footnotetext{Hossein Kavianirad, Satoshi Endo, and Sandra Hirche are with the Chair of Information-oriented Control, TUM School of Computation, Information and Technology, Technical University of Munich, 80333 Munich, Germany (e-mails: hossein.kavianirad, s.endo, hirche@tum.de).}
\footnotetext{Davide Astarita, Lorenzo Amato, and Emilio Trigili are with the BioRobotics Institute, Scuola Superiore Sant’Anna, 56025 Pontedera, Italy, and the Department of Excellence in Robotics and AI, Scuola Superiore Sant’Anna, 56127 Pisa, Italy (e-mails: davide.astarita, lorenzo.amato, emilio.trigili@santannapisa.it).}
\endgroup

\begin{abstract}
%
Hybrid assistive systems that integrate functional electrical stimulation (FES) and robotic exoskeletons offer a promising approach for neurorehabilitation. However, control of these systems remains challenging due to actuator redundancy and heterogeneous assistive device constraints. This paper introduces a novel cooperative control architecture based on dynamic allocation to address actuator redundancy in a hybrid FES-exoskeleton system. The proposed approach employs a modular control allocator that redistributes required control torques between FES and exoskeleton actuators in real time, accounting for device-specific limitations and user preferences (e.g., prioritizing one assistive device over another). Within this framework, the high-level controller determines the total assistance level, while the allocator dynamically distributes control effort based on these assistive device-specific considerations. Simulation results and experimental validation demonstrate the method's effectiveness in resolving actuator redundancy in the FES-exoskeleton system while reflecting actuator constraints, indicating its potential for deployment in clinical studies to assess patient acceptance and clinical efficacy.

\end{abstract}


\section{Introduction}
\revisionStatus{3}
Hybrid exoskeletons integrate functional electrical stimulation (FES)~\cite{rushton2003functional, lynch2008functional, previdi2003design, marquez2020functional, brend2015multiple} and exoskeletons~\cite{gull2020review, ochieze2023wearable, pan2023self, penna2024muscle} to enhance neuro-rehabilitation by leveraging the advantages of both technologies. FES actively recruits muscle fibers and stimulates the neuromuscular system, while the exoskeleton applies external torque to support the limb motion. Although there is evidence of FES-induced neuroplasticity following prolonged use~\cite{popovic2004therapy}, it remains challenging to realize precise goal-oriented assistance due to nonlinear and complex neuromuscular dynamics in response to stimulation~\cite{sena2022gap, kavianirad2024toward}. Moreover, FES accelerates muscle fatigue, which degrades FES-induced torque generation and consequently deteriorates control performance of the system~\cite{Dunkelberger2020ARO}. The active exoskeleton, in contrast, provides accurate and smooth joint motion, but the motions are externally guided and do not directly activate muscles~\cite{pan2022nesm}. By combining these technologies, hybrid systems promise to benefit from the strengths of both while compensating for their respective limitations~\cite{Kirsch18, Wolf17, hohler2024efficacy, cousin2021switched, molazadeh2021iterative, casas2023switched}: FES actively contracts muscle tissue to assist individuals with impaired motor function, while the exoskeleton provides motion and support to the affected limb through external forces that compensate for the limitations of FES-induced torque~\cite{stewart2017review, kavianirad2024toward}. Recent studies have demonstrated the potential of FES-exoskeleton systems to assist goal-directed movements in individuals with motor impairments, particularly in post-stroke rehabilitation~\cite{stewart2017review, romero2019design, dunkelberger2020review, dunkelberger2023hybrid}.

In a hybrid FES-exoskeleton, both assistive devices contribute to the net torque at the same joint, resulting in an inherently overactuated system: the exoskeleton produces torque mechanically and FES induces torque via muscle activation on the same joint.
Actuator redundancy provides additional degrees of freedom for control, allowing the consideration of factors such as input saturation and rate constraints, actuator fault tolerance, and secondary objectives~\cite{johansen2013control}. 
While the primary objective is typically to minimize the tracking error, secondary objectives are often selected from an operational perspective, such as reducing power consumption or prioritizing one actuator over the other for rehabilitation purposes or fatigue recovery. The control challenge and aim of this work is to design a control allocation that appropriately distributes the control effort among the actuators while ensuring their desired coordination~\cite {tregouet2024input, passenbrunner2016optimality}. Furthermore, since the dynamical response of the musculoskeletal system to assistive devices is complex, it is crucial to account for the dynamic behaviors and limitations of respective actuators when designing control allocation. 

State-of-the-art approaches for distributing control in over-actuated systems, while considering actuator dynamics and constraints, include optimal control and dynamic allocation methods~\cite{galeani2008magnitude, zaccarian2009dynamic, harkegaard2005resolving, johansen2013control, cocetti2016dynamic, romero2019design, dunkelberger2023hybrid}. Optimal control methods simultaneously determine the overall level of actuation and distribute redundant actuation by optimizing a cost function that incorporates system performance and redundant control inputs~\cite{romero2019design,dunkelberger2023hybrid}.

Dynamic allocation~\cite{zaccarian2009dynamic}, on the other hand, offers a more practical solution for addressing redundancy in over-actuated systems in a more modular approach, presenting several advantages over constant allocation, which uses a predefined fixed ratio for distributing control effort among actuators, and optimal control~\cite{zaccarian2009dynamic, harkegaard2005resolving, Sadien}. Dynamic allocation of control inputs refers to dynamically selecting the most appropriate control effort distribution from a set of feasible distributions that maintain theoretically identical system states (strong input redundancy) or at least steady-state system output (weak input redundancy)~\cite{zaccarian2009dynamic, johansen2013control, cocetti2016dynamic}. Control is allocated based on actuator constraints (e.g., torque magnitude, rate of change saturation), user preferences (e.g., favoring one actuator over another), and performance criteria (e.g., minimizing tracking error)~\cite{galeani2008magnitude}. In addition to being able to incorporate these design considerations, dynamic allocation is computationally efficient and easy to tune compared to optimal control~\cite{harkegaard2005resolving}.
Another key advantage is its modularity, allowing for an independent, low-complexity design of the high-level control and allocation scheme. This enables the allocation scheme to work as a plugin with minimal modifications to the existing high- and low-level control architecture, thereby facilitating adaptation to different user neuromuscular characteristics. 
%
%
Dynamic allocation is able to redistribute control effort computed by the high-level controller among actuators in real-time, based on design considerations, with minimal impact—ideally no effect in theory—on the states or at least the steady-state output of the system~\cite{zaccarian2009dynamic, harkegaard2005resolving, kreiss2021input, tregouet2024input}.

Solving input redundancy in hybrid FES-exoskeleton systems presents several key challenges that cooperative control needs to address. 
First, the control distribution needs to automatically adapt to heterogeneous, state- and time-dependent actuator constraints without requiring a redesign of the control system or manual parameter tuning, which is often needed in optimization-based methods.
Second, the human-in-the-loop nature of the control problem necessitates respecting the safety and comfort of the user, which imposes additional limitations on actuator usage and the attainable set of each actuator~\cite{kavianirad2024toward}.
Third, actuator usage priorities determined by system objectives or user preferences (e.g., prioritizing FES for rehabilitation uptake) must be incorporated into the allocation strategy.

Only very few studies have considered input redundancy in hybrid FES-exoskeleton systems~\cite{ha2015approach, romero2019design, dunkelberger2023hybrid, dalla2024hybrid}. However, even in these studies, the main challenge of solving actuator redundancy in FES-exoskeleton systems remains largely unresolved. For instance, the authors in \cite{dalla2024hybrid} assume a linear relationship between stimulation charge and generated torque, and their proposed cooperative control relies on a predefined, constant allocation ratio that typically fails to capture the time- and state-dependent constraints of the actuators. Other works have considered optimization-based strategies: a muscle model with inverse dynamics optimization is used in~\cite{romero2019design}. Model predictive control (MPC) is proposed in~\cite{dunkelberger2023hybrid}, where the cost function comprises trajectory tracking error, the control input magnitude, and the rate of change of the control input. However, these optimization-based methods are computationally expensive, and demand extensive parameter tuning—often through trial and error—as well as potential adjustment of the entire control architecture to reflect the new constraints. 

In this research, we propose a novel cooperative control framework based on dynamic allocation, which offers modularity and adaptability~\cite{harkegaard2005resolving, johansen2013control, argha2019control}, and functions as a plugin that can seamlessly adjust to actuator constraints without requiring adjustment in the high-level or entire control architecture. Moreover, it automatically determines and updates the allocation based on design objectives without relying on heuristic or user-defined parameters. This highlights the potential of dynamic allocation methods in addressing the challenges of input redundancy in hybrid FES-exoskeletons, enabling effective distribution of assistance while respecting human-in-the-loop limitations, actuator constraints, user preferences, and overall control objectives.

\subsection{Contribution}
The main contribution of this article is the design of a dynamic allocation-based cooperative control for a hybrid FES-exoskeleton. 
Thus, the proposed architecture enables real-time distribution of control efforts between functional electrical stimulation (FES) and exoskeleton, while accounting for actuator limitations, user preferences, and control objectives.
%
%
%
To achieve this, we introduce a control architecture consisting of a high-level control that determines the total assistive torque and a modular dynamic allocation scheme that distributes the net torque among the assistive devices.
%
The framework introduces a governing equation for torque distribution dynamics, providing interpretable allocation parameters that can adapt to the aforementioned design considerations.
The proposed methods are validated in numerical simulations and user testing, where the dynamic allocation adapts the distribution ratio online based on actuator constraints. 

In our preliminary work~\cite{kavianirad2023model}, we proposed effort allocation control in a hybrid FES-exoskeleton system; however, it employed a constant allocation ratio. Constant allocation relies on a predetermined ratio and is not capable of addressing state- and time-dependent actuator characteristics or of adapting to control objectives such as prioritizing one actuator over the other for rehabilitation purposes or fatigue recovery.

\subsection{Paper Organization}

The article is organized as follows. Section~\ref{sec:Hybrid Control Scheme} describes the hybrid control architecture addressing both high- and low-level control of the system. This section is further divided into three parts:
Section~\ref{sec:shared_cooperative_control} introduces the shared and cooperative control strategy that addresses actuator redundancy through dynamic allocation; Section~\ref{sec:fes_control} describes the FES torque model and control; and Section~\ref{sec:exo_control} presents the exoskeleton control strategy.
Section~\ref{sec:Experimental Evaluation} elaborates on the experimental evaluation, and the results are presented in Section~\ref{sec:Results}. Finally, a conclusion is provided in Section~\ref{sec:Conclusion}.

\section{Hybrid Control Scheme} \label{sec:Hybrid Control Scheme}
The hybrid assistive technology integrating FES and powered exoskeleton, along with an overview of the control architecture, is depicted in Fig.~\ref{fig:High-level_control_with_user}. We propose this control architecture to address the control allocation challenges introduced by actuator redundancy.
%

\subsection{Shared and Cooperative Control}\label{sec:shared_cooperative_control}
\revisionStatus{1}

\begin{figure*}[ht]
     \centering
\includegraphics[width=0.7\textwidth,trim={0.4cm 6.40cm 5.55cm 18.45cm},clip]{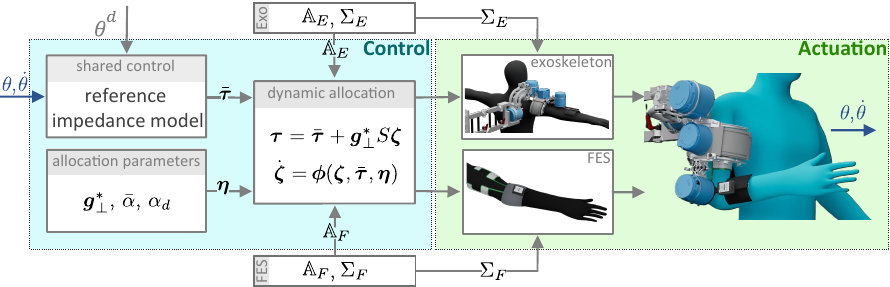}
     \caption{\textbf{Control architecture of the hybrid FES-exoskeleton system.} 
     The proposed dynamic allocation scheme distributes the control torque, determined by the shared control, among the actuators. The adaptation of the cooperative control coefficient is based on the attainable set of both assistive devices. The attainable set of FES-induced control torque, \(\mathbb{A}_F\), and attainable set of exoskeleton control torque, \(\mathbb{A}_E\), consists of constraints of two actuators. 
     FES-torque model, \(\Sigma_F\), and attainable sets of FES, \(\mathbb{A}_F\), learned from user data, are used in the low-level FES control and dynamic allocation, respectively.
     }
     \label{fig:High-level_control_with_user}
\end{figure*}

We propose a torque-based control architecture to achieve modularity in hybrid FES-exoskeletons with redundant actuation; volitional, FES, and exoskeleton torque. This architecture enables modularity by leveraging the superposition of torques generated by actuators at the joint level. In contrast, kinematic control does not support superposition of actuator contributions in position or velocity space, limiting modular coordination of multiple actuators at the joint level. Fig.~\ref{fig:High-level_control_with_user} depicts the high-level controller incorporating shared and cooperative control strategies. Shared control determines the total assistance by taking into account volitional human effort and provides robustness against model and environmental uncertainties, while cooperative control addresses the distribution of assistance among the assistive technologies.




We employ a shared control framework based on a reference impedance model~\cite{sup2008design, chen2014fes, gehlhar2023review, kavianirad2023model}, which computes the total control effort required from the assistive devices to minimize tracking error. Dynamic allocation is the proposed cooperative control approach for dealing with input redundancy in this research. Mathematically, in dynamic allocation, an undetermined system of equations known as the allocator dynamics, which is often subject to additional constraints such as actuator limitations and secondary control objectives, is solved.






\subsubsection{Dynamic Allocation in Redundant System}
Consider an affine nonlinear system \(\Sigma_H\) of the form
\begin{subequations} \label{eq:eq_nonlin}
\begin{align}
    &\dot{\boldsymbol{x}} = \boldsymbol{f}(\boldsymbol{x}) + \boldsymbol{g}(\boldsymbol{x})\boldsymbol{u}, \label{eq:eq_nonlin_m} \\
    &\boldsymbol{y} = \boldsymbol{c}(\boldsymbol{x}) + \boldsymbol{d}(\boldsymbol{x})\boldsymbol{u}, \label{eq:eq_nonlin_b}
\end{align}
\end{subequations}
\noindent where \(\boldsymbol{x} \in \mathbb{R}^n\) is the plant state, \(\boldsymbol{u} \in \mathbb{R}^{n_u}\) is the plant input, and \(\boldsymbol{y} \in \mathbb{R}^{n_y}\) is the plant output, respectively. 

The system exhibits strong input redundancy if there exist admissible variations in \(\boldsymbol{u}\) that do not affect the system’s states and output. This condition holds when the augmented control matrix \(\boldsymbol{g}^*(\boldsymbol{x}) = [\boldsymbol{g}(\boldsymbol{x}), \boldsymbol{d}(\boldsymbol{x})]^\top \) is rank-deficient, i.e. \(\mathrm{rank}([\boldsymbol{g}(\boldsymbol{x}), \boldsymbol{d}(\boldsymbol{x})]^\top ) < n_u\).


Let us define 
\begin{subequations} \label{eq:g_ortho}
\begin{align}
    &\text{Im}(\boldsymbol{g}^*_{\perp}) = \text{Ker}([\boldsymbol{g}, \boldsymbol{d}]^\top ), \label{eq:g_ortho_a} \\
    &\boldsymbol{u} = \bar{\boldsymbol{u}} + \boldsymbol{g}^*_{\perp} \boldsymbol{\zeta}, \label{eq:g_ortho_b}
\end{align}
\end{subequations}
where \(\bar{\boldsymbol{u}} \in \mathbb{R}^{m \times 1}\) is the nominal control input and \(\boldsymbol{\zeta} \in \mathbb{R}^{l \times 1}\) (\(l = \mathrm{rank}(\boldsymbol{g}^*_{\perp})\)) is an arbitrary signal. 


The redistribution vector \(\boldsymbol{g}^*_{\perp} \boldsymbol{\zeta}\) is invisible to high-level controllers~\cite{zaccarian2009dynamic}; in other words, perturbing the control input by any vector of the form \(\boldsymbol{g}^*_{\perp} \boldsymbol{\zeta}\), where \(\boldsymbol{\zeta}\) is arbitrary, does not affect the system state \(\boldsymbol{x}\) and the output \(\boldsymbol{y}\). Mathematically, for \(\Sigma_H\)
\begin{equation}\label{eq:IR_mathmatically}
\forall\, (\bar{\boldsymbol{u}}, \bar{\boldsymbol{x}}, \bar{\boldsymbol{y}}),\ (\boldsymbol{u}, \boldsymbol{x}, \boldsymbol{y}) \in Q(\boldsymbol{x}_0): \bar{\boldsymbol{x}} = \boldsymbol{x},\ \bar{\boldsymbol{y}} = \boldsymbol{y},
\end{equation}
\noindent where \(\boldsymbol{x}_0 \in \mathbb{R}^n\) is a given initial condition, and the set of all triples \((\boldsymbol{u}, \boldsymbol{x}, \boldsymbol{y})\) compatible with \(\boldsymbol{x}_0\) is denoted by \(Q(\boldsymbol{x}_0)\).

\begin{remark}[Redundancy as an Additional Degree of Freedom]
Redundancy in an over-actuated system (mathematically \(\mathrm{rank}(\boldsymbol{g}^*) < n_u\)) allows for altering the control input vector (i.e., redistributing control effort) without affecting the system output. This provides additional degrees of freedom, enabling the incorporation of secondary objectives. In hybrid FES-exoskeleton systems, we aim to leverage this redundancy to accommodate the capabilities and constraints of assistive devices, as well as their usage priorities, in control allocation. \remarkend
\end{remark}


\subsubsection{Input Redundancy in Hybrid FES-Exoskeleton}\label{section: Input redundancy in hybrid exoskeleton}


In a hybrid FES-exoskeleton system, the plant state is represented as \(\boldsymbol{x} = [\boldsymbol{\theta}, \dot{\boldsymbol{\theta}}]^\top \), where \(\boldsymbol{\theta} \in \mathbb{R}^n\) denotes the vector of all joint angles. 
The plant input is defined as \(\boldsymbol{u} = [\boldsymbol{\tau}_1^\top , \boldsymbol{\tau}_2^\top , \dots, \boldsymbol{\tau}_n^\top ]^\top \),  where each \(\boldsymbol{\tau}_j \in \mathbb{R}^3\) represents the torque input vector at joint \(j\), comprising FES-induced torque on the flexor and extensor muscles, as well as exoskeleton-generated torque. For simplicity and without loss of generality, we omit the subscript \(j\) when discussing a single joint, using the notation \(\square\triangleq \square_j\)
\begin{equation} \label{eq:torque_vec}
    \boldsymbol{\tau} = [{\tau}^{F_f}, {\tau}^{F_e}, {\tau}^{E}]^\top, 
\end{equation}
where \(\tau^{F_f}\), \(\tau^{F_e}\), and \(\tau^E\) represent the flexor FES-induced torque, extensor FES-induced torque, and exoskeleton torque at joint \(j\), respectively. 

Although the flexor and extensor muscle groups for a given joint both generate torque about the same joint, they are distinct actuators with potentially different physiological characteristics, constraints, and dynamic responses to stimulation. Therefore, FES stimulation of flexor and extensor muscles is treated as separate actuators. A typical example is elbow flexion and extension, where a pair of antagonistic muscles actuates a single degree of freedom.  

We approximate the muscle co-contraction torque induced by FES~\cite{bo2016fes}, \({\tau}^{C}\), on joint \(j\) as the minimum absolute torque of the antagonistic pair, capturing the stabilizing effect of simultaneous activation
\begin{equation} \label{eq:torque_cc}
\begin{split}
    &{\tau}^{C} = \min(|\tau^{F_f}|, |\tau^{F_e}|). 
\end{split}
\end{equation}
This co-contraction does not contribute to the net FES-induced torque but modulates the impedance of the joint. 

Therefore, we can decompose flexor and extensor FES torque as follows
\begin{equation} \label{eq:effective_muscle_induced}
\begin{split}
&[\tau^{F_f},\tau^{F_e}]^\top  = [h(\tau^{F}),1-h(\tau^{F})]^\top \tau^{F} + [1,-1]^\top {\tau}^{C}, \\
\end{split}
\end{equation}
where \(h(.)=1/2(1+sign(.))\) is the Heaviside step function and \(\tau^{{F}}\) indicate the FES induced net torque on joint \(j\). 
This decomposition allows us to rewrite the torque at joint \(j\) \eqref{eq:torque_vec}, as follows
\begin{equation} \label{eq:torque_vec_2}
\begin{split}
    &\boldsymbol{\tau} = [{\alpha}\boldsymbol{\sigma}, 1 - {\alpha}]^\top {\tau}^{N} +  \boldsymbol{c}{\tau}^{C}, \\
    &\boldsymbol{\sigma}({\tau}^{N}) = [h({\tau}^{N}), 1 - h({\tau}^{N})], \\
    &\boldsymbol{c} = [1, -1, 0]^\top , \\
\end{split}
\end{equation}
where \({\tau}^{N} =\boldsymbol{1}^\top \boldsymbol{\tau}  \in \mathbb{R}\) represents net control torque at the redundantly actuated joint \(j\) and \({\alpha}\) represents the cooperative gain, which defines the contribution of different sources of torque in inducing net joint torque. \({{\alpha}}\) is defined as the ratio of the FES-induced torque to the net control torque
\begin{equation} \label{eq:alpha}
\begin{split}
    &{\alpha}= \frac{\tau^{{F}}}{{\tau}^{N}}.
\end{split}
\end{equation}
Note that the ratio is undefined when \( {\tau}^{N}=0 \); however, for implementation purposes, we retain the last valid value of \({\alpha}\) in such cases.

Furthermore, \(\boldsymbol{\sigma}\) can be interpreted as a FES-induced muscle torque distributor that allocates the FES control effort between the flexor and extensor muscles.



The nominal control input in~\eqref{eq:g_ortho}, denoted by \(\bar{\boldsymbol{u}}=[\bar{\boldsymbol{\tau}}_1^\top , \bar{\boldsymbol{\tau}}_2^\top , \dots, \bar{\boldsymbol{\tau}}_n^\top ]^\top \) in the FES-exoskeleton system refers to any control input that satisfies the desired system behavior. This can be derived from high-level control strategies such as impedance or optimal control.
Similar to \eqref{eq:torque_vec_2}, we rewrite the nominal control input at joint \(j\) as
\begin{equation} \label{eq:torque_nom}
\begin{split}
    &\bar{\boldsymbol{\tau}} = [{\bar{\alpha}}\boldsymbol{\sigma}, 1 - {\bar{\alpha}}]^\top \bar{{\tau}}^{N} +  \boldsymbol{c}{\tau}^{C}, \\
\end{split}
\end{equation}
where \(\bar{{\tau}}^{N} =\boldsymbol{1}^\top \bar{\boldsymbol{\tau}}  \in \mathbb{R}\) represents nominal net control torque at joint \(j\).
Similar to \eqref{eq:alpha}, the contribution of different actuators in the nominal control input \(\bar{\boldsymbol{\tau}}\) can also be parameterized by the nominal cooperative gain \({\bar{\alpha}}\)
\begin{equation} \label{eq:alpha_bar}
\begin{split}
    &\bar{\alpha}= \frac{\bar{\tau}^{{F}}}{\bar{{\tau}}^{N}},
\end{split}
\end{equation} 
\noindent where \( \bar{\tau}^{{F}} \) is nominal FES-induced torque.


\begin{remark}[Nominal Control Input]
When the nominal control input is determined by impedance control, selecting \(\bar{\alpha} = 0\), though not the only possible choice, is an intuitive choice. This implies that, prior to any redistribution of control effort, the entire desired torque is assumed to be provided by the stronger actuator, namely the exoskeleton. Without loss of generality, the proposed framework also accommodates high-level control strategies that already account for individual actuator contributions—such as constant allocation or optimal control. In such cases, the modular dynamic allocation can still be applied to redistribute the already allocated control torque among the actuators. 
\end{remark}

\begin{remark}[FES as One or Two Actuators]
Where the distinction between flexor and extensor muscles is not possible, or an exclusive mapping unavailable, FES can be treated as a single actuator by redefining the control input in~\eqref{eq:torque_vec} as \(\boldsymbol{\tau} = [{\tau}^{F}, {\tau}^{E}]^\top\). In this case, there is no need to define muscle co-contraction \({\tau}^{C}\) in~\eqref{eq:torque_cc}, and we can directly use~\eqref{eq:alpha}. \remarkend
\end{remark}

\subsubsection{Dynamic Allocation in Hybrid FES-Exoskeleton}\label{section: Dynamic allocation in hybrid exoskeleton}
Let us define the following set of equations as the governing equations of the dynamic allocator in the FES-exoskeleton system
\begin{subequations} \label{eq:allocator_structure}
\begin{align}
    &\boldsymbol{\tau} = \bar{\boldsymbol{\tau}} + \boldsymbol{g}^*_{\perp} S \boldsymbol{\zeta}, \label{eq:allocator_structure_a} \\
    &\dot{\boldsymbol{\zeta}} = \boldsymbol{\phi}(\boldsymbol{\zeta}, \bar{\boldsymbol{\tau}}, \boldsymbol{\eta}), \label{eq:allocator_structure_b} 
\end{align}
\end{subequations} 
\noindent where the dynamics in~\eqref{eq:allocator_structure_b} are adopted from the allocation framework in~\cite{cocetti2016dynamic, cocetti2018linear}. The torque input vector \(\boldsymbol{\tau}\) is composed of a nominal input \(\bar{\boldsymbol{\tau}}\) and a null-space contribution given by \(\boldsymbol{g}^*_{\perp} S \boldsymbol{\zeta}\), where \(\boldsymbol{g}^*_{\perp}\) denotes the basis of the null space of the augmented control matrix of the hybrid FES-exoskeleton 
\begin{equation} \label{eq:g_orth}
\begin{split}
    &\boldsymbol{g}^*_{\perp} =
\begin{bmatrix}
1 & 1 \\
0 & -1 \\
-1 & 0
\end{bmatrix},
\end{split}
\end{equation} 
\noindent where \(l=\mathrm{rank}(\boldsymbol{g}^*_{\perp}) = 2\). \(S \in \mathbb{R}^2\) is an arbitrary matrix and the vector \(\boldsymbol{\zeta} = [{\zeta}_{1}, {\zeta}_{2}]^\top \in \mathbb{R}^2\) represents the dynamic input allocator. The parameter set \(\boldsymbol{\eta}\) includes all relevant information affecting the allocator dynamics, such as user preferences (e.g., prioritizing one actuator over another), alongside nominal cooperative gain \(\bar{\alpha}\), basis of the null space of augmented control matrix \(\boldsymbol{g}^*_{\perp}\), and FES-induced torque distributor vector~\(\boldsymbol{\sigma}\). 


\begin{assumption} \label{assump:internal_stability}
The allocation dynamics in~\eqref{eq:allocator_structure_b} is input-to-state stable with respect to \(\bar{\boldsymbol{\tau}}(t)\), i.e., for any bounded nominal input \(\bar{\boldsymbol{\tau}}(t)\), the allocator state \(\boldsymbol{\zeta}(t)\) remains bounded for all $t \geq 0$. 
\end{assumption}


Mathematically, \(\boldsymbol{\zeta}(t)\) in~\eqref{eq:allocator_structure_b} remains bounded for bounded inputs \(\bar{\boldsymbol{\tau}}(t)\) if there exist functions \( \mu \in \mathcal{K_\infty} \) and \( \delta \in \mathcal{KL} \) such that, for every initial state \( \boldsymbol{\zeta}(0) \) and every input \(\bar{\boldsymbol{\tau}}(t)\), the corresponding solution of~\eqref{eq:allocator_structure_b} satisfies the inequality
\begin{equation} \label{eq:ISS_def}
    \|\boldsymbol{\zeta}(t)\| \leq \delta(\|\boldsymbol{\zeta}(0)\|, t) + \mu\left(\|\bar{\boldsymbol{\tau}}\|_{[0,t]}\right) , \forall t \geq 0
\end{equation}
where \(\|.\|\) denotes Euclidean norm, and \( \|\bar{\boldsymbol{\tau}}\|_{[0,t]} := \mathop{\mathrm{ess\,sup}} \{ \|\bar{\boldsymbol{\tau}}(p)\| : p \in [0, t] \} \) (supremum norm on \([0,t]\) except for a set of measure zero)~\cite{sontag1998comments}~\cite[Appendix A.6]{liberzon2003switching}.

\begin{remark}
Given condition~\eqref{eq:g_ortho}, input redundancy~\eqref{eq:IR_mathmatically} holds for any arbitrary signal \(\boldsymbol{\zeta}\). However, \(\boldsymbol{\zeta}\) modifies the individual actuator torques and, therefore, \(\boldsymbol{\zeta}(t)\) must remain bounded for bounded inputs. As a counterexample, \(\dot{{\zeta}}_1 = -{\zeta}_1 + \bar{{\tau}}_1 {\zeta}_1\) is not suitable candidate for dynamic allocations since \({\zeta}_1(t)\) becomes unbounded even for bounded inputs \(1<\bar{{\tau}}_1 < \bar{{\tau}}^{max}_1\), which can result in actuator commands that violate actuation feasibility.
\end{remark}

\begin{theorem} \label{theorem:invisibility}
Suppose Assumption~\ref{assump:internal_stability} holds for the hybrid system \(\Sigma_H\)~\eqref{eq:eq_nonlin} with the allocation dynamics~\eqref{eq:allocator_structure}. Then, the redistribution term \(\boldsymbol{g}^*_{\perp}S\boldsymbol{\zeta}\) is invisible to \(\Sigma_H\)
\begin{subequations} \label{eq:IR_mathmatically_2}
\begin{align}
 &\boldsymbol{x}(t,x_0, \bar{\boldsymbol{\tau}}) = \boldsymbol{x}(t,x_0, \bar{\boldsymbol{\tau}}+ \boldsymbol{g}^*_{\perp} S \boldsymbol{\zeta}),\\
 &\boldsymbol{y}(t,x_0, \bar{\boldsymbol{\tau}}) = \boldsymbol{y}(t,x_0, \bar{\boldsymbol{\tau}}+ \boldsymbol{g}^*_{\perp} S \boldsymbol{\zeta}), \quad \forall t \geq 0,
\end{align}
\end{subequations}
\end{theorem}

\textbf{Proof:}  
The proof is given in Appendix~\ref{sec:Apendix_A}. \theoremend




\begin{remark}[Attainable Sets Represent Actuator Constraints] \label{lemma:Attainable_Set_1}
The low-level control torques, \({\tau}^F\) and \({\tau}^E\), as a function of nominal control torque \(\bar{\boldsymbol{\tau}}\) and dynamic input allocator \(\boldsymbol{\zeta}\), must lie within the attainable sets of the FES-induced torque, \(\mathbb{A}_F\), and the exoskeleton torque, \(\mathbb{A}_E\), respectively. This can be expressed as follows
\begin{subequations} \label{eq:actuator_constraints}
\begin{align}
 &{\tau}^F \in \mathbb{A}_F ({\tau}^{F_f} \in \mathbb{A}_{F_f},{\tau}^{F_e} \in \mathbb{A}_{F_e}), \\
 &{\tau}^E \in \mathbb{A}_E,
\end{align}
\end{subequations}
\noindent where \({\tau}^{F_f}\) and \({\tau}^{F_e}\) are  flexor and extensor components of \({\tau}^{F}\). These sets account for the attainable torque of both assistive devices while incorporating their magnitude and bandwidth constraints. Note that bandwidth limits are associated with the actuator dynamics (e.g., rate and acceleration constraints)~\cite{braun2013robots, neilson1972speed, winters2007analysis}.  

Fig.~\ref{fig:Attainable_Set_1} illustrates the magnitude constraints of these two assistive devices on the biceps side. FES is known not to provide accurate torque by itself, leading to a discrepancy in torque accuracy between FES and the exoskeleton. This discrepancy is reflected in the attainable set shown in Fig.~\ref{fig:Attainable_Set_1}.
\begin{figure}[!ht]
     \centering
     \vspace*{0mm}
\includegraphics[width=0.48\textwidth,clip,trim={0.05cm 22.25cm 10.6cm 0.43cm}]{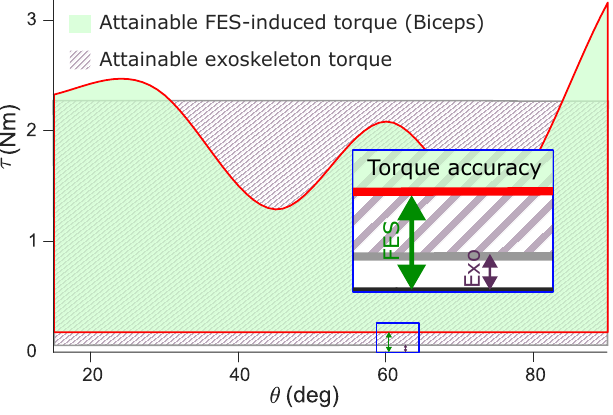}
      \vspace*{-0mm}
     \caption{\textbf{Attainable set of FES-induced control torque and exoskeleton torque.} Although bandwidth limitations are part of attainable sets \(\mathbb{A}_F\) and \(\mathbb{A}_E\), this figure only depicts the magnitude constraints of two assistive devices. Note that \(\mathbb{A}_F\) on this figure represents the magnitude limitations of FES-induced biceps torque derived from experimental data.}
     \vspace*{-0mm}
     \label{fig:Attainable_Set_1}
\end{figure}
\remarkend
\end{remark}

\begin{remark}[User Safety and Comfort Limitations] \label{lemma:bandwidth}
In a hybrid FES-exoskeleton, the exoskeleton must limit its maximum torque to ensure user safety, and the FES, which delivers electrical stimulation directly to muscles, has to respect comfort and safety constraints by limiting both the magnitude and rate of stimulation intensity. These input limitations result in torque constraints (i.e., in magnitude and bandwidth) for both the exoskeleton and FES subsystems in open- or closed-loop low-level control. Such constraints restrict the attainable sets of each actuator and must be considered in cooperative control schemes.
In this work, constraints of the assistive devices are determined based on the identification of the system. In particular, FES-induced torque constraints are identified by evaluating the system’s response to the maximum safe and comfortable stimulation intensity. By incorporating these constraints,  the low-level desired torque determined by dynamic allocation remains within the attainable set of each actuator, while respecting user safety and comfort. Therefore, in human-centered systems, constraints imposed by user safety and comfort significantly impact the control design. \remarkend 
\end{remark}

\begin{remark}[Hybrid System Saturation and Anti-Windup] \label{lemma:Attainable_Set_1general}
High-level control should account for the cumulative attainable sets of the two actuators (\(\mathbb{A}_F \cup \mathbb{A}_E\)). Otherwise, it may lead to control saturation, necessitating the implementation of an anti-windup mechanism~\cite{galeani2008magnitude, zaccarian2009dynamic}. \remarkend 
\end{remark}

Inspired by~\cite{zaccarian2009dynamic}, let define allocation dynamics, \(\dot{\boldsymbol{\zeta}}=\boldsymbol{\phi}(\cdot\)) in \eqref{eq:allocator_structure_b}, as follows 
\begin{equation} \label{eq:zeta_eq_2}
\begin{split}
    &\dot{\boldsymbol{\zeta}}=-KS{\boldsymbol{g}^*_{\perp}}^\top W\boldsymbol\tau,\\
\end{split}
\end{equation}
\noindent where \(K = \text{diag}([k_1, k_2]) \in \mathbb{R}^{2 \times 2}\) and \(W(t) = \text{diag}([w_1(t), w_2(t), w_3(t)]) \in \mathbb{R}^{3 \times 3}\) are positive definite diagonal matrices, and \(S = \mathrm{diag}(\boldsymbol{\sigma})\) is a diagonal matrix representing FES-induced muscle torque distributor. These are parameters of the allocation: \(K\) adjusts the convergence speed of the allocation dynamics, while \(W(t)\) modulates the ratio of allocation. The effects of these parameters on the allocation will be discussed in detail later.

\begin{theorem} \label{theorem2}
The allocation dynamics given by~\eqref{eq:zeta_eq_2} is input-to-state stable with respect to \(\bar{\boldsymbol{\tau}}(t)\), and the redistribution term \(\boldsymbol{g}^*_{\perp}S\boldsymbol{\zeta}\) in~\eqref{eq:allocator_structure_a} is invisible to \(\Sigma_H\)~\eqref{eq:eq_nonlin}. 
\end{theorem}

\textbf{Proof:} To show that the allocation dynamics \eqref{eq:zeta_eq_2} is input-to-state stable with respect to \(\bar{\boldsymbol{\tau}}(t)\), consider the positive definite Lyapunov function \(V(\boldsymbol{\zeta})=1/2\boldsymbol{\zeta}^\top K^{-1}\boldsymbol{\zeta}\). We can write
\begin{equation} \label{eq:d_v}
\begin{split}
    &\dot{V}({\boldsymbol{\zeta}})={{\boldsymbol{\zeta}}^\top K^{-1}\dot{\boldsymbol{\zeta}}} \\
    &\dot{V}({\boldsymbol{\zeta}})={-{\boldsymbol{\zeta}}^\top S{\boldsymbol{g}^*_{\perp}}^\top W\boldsymbol{\tau}} \\
\end{split}
\end{equation} 
\noindent given the definition of \( \tau \) in \eqref{eq:allocator_structure_a},
\begin{equation} \label{eq:d_v_2}
\begin{split}
    &\dot{V}({\boldsymbol{\zeta}})=-{\boldsymbol{\zeta}}^\top S{\boldsymbol{g}^*_{\perp}}^\top W\boldsymbol{g}^*_{\perp}S{\boldsymbol{\zeta}}-{\boldsymbol{\zeta}}^\top S{\boldsymbol{g}^*_{\perp}}^\top W\bar{\boldsymbol{\tau}}, \\
\end{split}
\end{equation} 

\noindent given that \(W(t)\) is positive definite by assumption and using the definition of \(\boldsymbol{g}^*_{\perp}\) in~\eqref{eq:g_orth}, \({\boldsymbol{g}^*_{\perp}}^\top W\boldsymbol{g}^*_{\perp}> 0\), and therefore \(S\boldsymbol{g}^{*^\top}_\perp W \boldsymbol{g}^*_\perp S \) is positive semi-definite. This implies that the first term of \(\dot{V}({\boldsymbol{\zeta}})\) in~\eqref{eq:d_v_2} is negative semi-definite 
\begin{equation} \label{eq:d_v_3}
\begin{split}
    &-{\boldsymbol{\zeta}}^\top S{\boldsymbol{g}^*_{\perp}}^\top W\boldsymbol{g}^*_{\perp}S{\boldsymbol{\zeta}} \leq 0, \\
\end{split}
\end{equation} 
\noindent moreover, we assume that the nominal control input \( \bar{\boldsymbol{\tau}}(t) \) is uniformly bounded for all \( t \geq 0 \), i.e., there exists a constant \( \bar{\tau}_{\max} > 0 \) such that \( \|\bar{\boldsymbol{\tau}}(t)\| \leq \bar{\tau}_{\max} \) for all \( t \). Then, making use of the Cauchy-Schwarz inequality, the second term in~\eqref{eq:d_v_2} can be upper bounded as

\begin{equation} \label{eq:d_v_4}
\begin{split}
    |\boldsymbol{\zeta}^\top S\boldsymbol{g}^{*^\top}_\perp W \bar{\boldsymbol{\tau}}| \leq  \|\boldsymbol{g}^{*^\top}_\perp W\|  \|S\boldsymbol{\zeta}\|\bar{\tau}_{\max}.
\end{split}
\end{equation} 
Substituting into~\eqref{eq:d_v_2} gives
\begin{equation} \label{eq:Vdot_iss_form}
    \dot{V}(\boldsymbol{\zeta}) \leq -\alpha \|S\boldsymbol{\zeta}\|^2 + \beta \|S\boldsymbol{\zeta}\|.
\end{equation}
where \( \alpha := \inf_{t \geq 0} (\lambda_{\min}({\boldsymbol{g}^*_{\perp}}^\top W(t)\boldsymbol{g}^*_{\perp})) \) and \( \beta := \sup_{t \geq 0} (\bar{\tau}_{\max}\|\boldsymbol{g}^{*^\top}_\perp W(t)\|) \). Here, \( \lambda_{\min}(\cdot) \) denotes the smallest nonzero eigenvalue of the matrix argument. Note that by assumption \(W(t)\) is a diagonal positive definite matrix (\(0 < w_{min} \leq w_i(t) \leq w_{max}, \quad \forall t \geq 0, \quad i=1,2, 3\)), which guarantees that \(\alpha,\beta > 0\) and \(\beta\) is finite.
Consequently, we obtain
\begin{equation} \label{eq:Vdot_iss_form_3}
    \dot{V}(\boldsymbol{\zeta})  <  0 \quad  \forall \, \|S\boldsymbol{\zeta}\| > \frac{\beta}{\alpha},
\end{equation}
hence, trajectories converge to 
\begin{equation} \label{eq:Vdot_iss_form_2}
    \Omega = \{(\zeta_1,\zeta_2) \; | \; \|S\boldsymbol{\zeta}\| \leq \frac{\beta}{\alpha}\},
\end{equation}

Now, given the definition \( S = \mathrm{diag}(\boldsymbol{\sigma}) \), see~\eqref{eq:torque_vec_2}, for \( \tau^N \geq 0 \), we have \( \|S \boldsymbol{\zeta}\| = |\zeta_1| \), and from~\eqref{eq:zeta_eq_2}, it follows that \( \dot{\zeta}_2 = 0 \), meaning \( \zeta_2 \) remains constant. Similarly, for \( \tau^N < 0 \), we obtain \( \|S \boldsymbol{\zeta}\| = |\zeta_2| \), and again from~\eqref{eq:zeta_eq_2}, \( \dot{\zeta}_1 = 0 \), so \( \zeta_1 \) remains constant. 
This guarantees that all signals within the allocation dynamics remain bounded for any bounded external input $\bar{\boldsymbol{\tau}}$, thereby ensuring that the allocation dynamics is input-to-state stable with respect to \(\bar{\boldsymbol{\tau}}\) (see Appendix~\ref{sec:Apendix_B} for more details). Consequently, the redistribution term \(\boldsymbol{g}^*_{\perp}S\boldsymbol{\zeta}\) in~\eqref{eq:allocator_structure_a} is invisible to \(\Sigma_H\)~\eqref{eq:eq_nonlin}.

Moreover, the matrices \(K\) and \(W(t)\) are chosen in a way to comply with two other conditions in \eqref{eq:actuator_constraints}, namely, \({\tau}^F \in \mathbb{A}_F\) and \({\tau}^E \in \mathbb{A}_E\).  \theoremend

\begin{remark}[Switching Allocation Dynamics]  \label{re:Switching Allocation Dynamics}
The allocation dynamics defined in~\eqref{eq:zeta_eq_2} represents a time-dependent switching system~\cite{liberzon2003switching}, where the switching signal is determined by the sign of \(\tau^N(t)\) as defined by \(S\). The rigorous proof of input-to-state stability of this switching allocation dynamics with signal bound derivation is established in Appendix~\ref{sec:Apendix_B}.
\remarkend
\end{remark}


\begin{remark}[Time-Varying Constraints]  \label{time-varying_w}
Time-varying \(W(t)\) allows for adaptation to dynamic constraints such as the time-varying magnitude saturation of FES-induced torque resulting from FES-induced muscle fatigue. \remarkend 
\end{remark}

Combining \eqref{eq:allocator_structure_a} and \eqref{eq:zeta_eq_2}, the allocator dynamics can be rewritten as 
\begin{equation} \label{eq:zeta_eq2}
\begin{split}
    \dot{\zeta}_i ={}& -k_i\sigma_i(w_i+w_3)\zeta_i \\
    &+ k_i\sigma_i\bigl(w_3-(w_i+w_3)\bar{\alpha}\bigr)\bar{\tau}^{N}, \quad i=1,2
\end{split}
\end{equation}
where \(\sigma_i\) denotes the \(i\)-th component of \(\boldsymbol{\sigma}\) in~\eqref{eq:torque_vec_2}.
Then, let define the steady-state cooperative gain and modified \(k_i\) denoted by \({\alpha}^s_i\) and \({k'_i}\) respectively, as follows 
\begin{equation} \label{eq:cooperative_gain}
\begin{split}
    &{\alpha}^s_i:=\frac{w_3}{w_i+w_3},\\
    &{k'}_i:=k_i\sigma_i({w_i+w_3}),\ i=1,2\\
\end{split}
\end{equation} 
given these new coefficients in \eqref{eq:cooperative_gain}, we can rewrite allocator dynamics \eqref{eq:zeta_eq2} as follows
\begin{equation} \label{eq:zeta_eq3}
\begin{split}
    &\dot{\zeta}_i=-k'_i{\zeta}_i+k'_i({\alpha}^s_i-\bar{\alpha})\bar{\tau}^{N},\ i=1,2\\
\end{split}
\end{equation} 

Note that by modifying \(k_1\), \(k_2\), \(w_1\), \(w_2\), and \(w_3\), or equivalently \(k'_1\), \(k'_2\), \({\alpha}^s_1\), and \({\alpha}^s_2\), one can adjust both the convergence speed and ratio of allocation.
Moreover, substituting the steady-state response of allocator dynamics, \eqref{eq:zeta_eq2} or \eqref{eq:zeta_eq3}, which is \({\zeta}^s_i=({\alpha}^s_i-\bar{\alpha})\bar{\tau}^{N}\), in \eqref{eq:allocator_structure_a}, we can write the steady-state FES-induced torque of flexor and extensor as
\begin{equation} \label{eq:tau_eq}
\begin{split}
    &\tau^{F_f, s} := \alpha^s_1 \bar{\tau}^{N}, \\
    &\tau^{F_e, s} := \alpha^s_2 \bar{\tau}^{N},
\end{split}
\end{equation}
\noindent this means that the steady-state cooperative gain, which determines the contribution of each actuator in the steady-state condition, is defined by the aforementioned gains.

The limitations of the actuators can be considered in allocator dynamics with the help of these gains which finally affects of convergence speed and ratio of allocation.

\textbf{Contribution of different actuators:}
For instance, increasing \(w_1\) (\(w_2\)), or equivalently decreasing \({\alpha}^s_1\) (\({\alpha}^s_2\)), reduces the contribution of the FES actuator in providing the desired flexion (extension) control torque. Similarly, increasing \(w_3\), or equivalently decreasing \({\alpha}^s_1\) and \({\alpha}^s_2\), requires the exoskeleton to deliver a higher share of the control effort.

\textbf{Allocation convergence speed:}
The parameter \(k'_i\) regulates the allocation convergence speed, determining how agile the system is in reallocating control efforts. 


\textbf{Actuator magnitude saturation:}
To account for actuator magnitude saturation, we consider \(W(\tau)\) in~\eqref{eq:zeta_eq_2} as described in~\cite{zaccarian2009dynamic}, in a way that \(w_i \rightarrow \infty\) or equivalently \({\alpha}^s_i \rightarrow 0\) and \({k'}_i \rightarrow \infty\) for FES saturation and \({\alpha}^s_i \rightarrow 1\) and \({k'}_i \rightarrow \infty\) for exoskeleton saturation when operating in the neighborhood of upper attainable torque of actuator \(i\) (\(m_i\))
\begin{equation} \label{eq:Actuator_saturation}
\begin{split}
    &\lim_{\tau_i \to m_i} w_i \to \infty, \\
\end{split}
\end{equation}

\textbf{Actuator rate saturation:}
The same method as \eqref{eq:Actuator_saturation} can be used to consider the actuator rate saturation~\cite{zaccarian2009dynamic}.

\textbf{Actuator bandwidth constraints:}
To account for actuator bandwidth, the convergence speed of the allocation dynamics is selected based on the bandwidth of the more constrained actuator (FES). This allows the incorporation of the lowest bandwidth into the allocation and prevents fast reallocation, and subsequently fast changes in desired FES torque, that would violate actuation feasibility. An alternative would be extension of actuator rate saturation~\cite{zaccarian2009dynamic} based on the method proposed in~\cite{braun2013robots}.


\begin{remark}[Advantages of Dynamic Allocation]  \label{Advantages of dynamic allocation}
It is worth highlighting again that dynamic allocation provides a modular and computationally efficient solution for real-time distribution of assistance between the exoskeleton and FES. Compared to optimization-based methods such as MPC, it requires only a few interpretable parameters—namely \(k'_1\) (\(k'_2\)) and \(\alpha^s_1\) (\(\alpha^s_2\))—which can be directly selected to reflect design objectives. The parameter \(\alpha^s_1\) (\(\alpha^s_2\)) specifies the desired contribution of different actuators, while \(k'_1\) (\(k'_2\)) determines the allocation convergence speed. Additionally, these parameters can be selected to directly encode desired steady-state cooperative gain and to account for actuator magnitude and bandwidth saturation.
\remarkend 
\end{remark}

\vspace{5pt}   

\textbf{Simulation:}
To provide an intuitive interpretation of the dynamic allocation scheme described in Section~\ref{section: Dynamic allocation in hybrid exoskeleton}, we present a simple example of dynamic allocation that incorporates the attainable sets and constraints of FES.

\textbf{Setup:}
We benchmark the proposed method from Section~\ref{section: Dynamic allocation in hybrid exoskeleton} on the modular cooperative control of a hybrid FES-exoskeleton. The allocator dynamics follows~\eqref{eq:allocator_structure_a} and ~\eqref{eq:zeta_eq_2} (equivalently \eqref{eq:allocator_structure_a} and \eqref{eq:zeta_eq3}), with the FES attainable set of the participant considered in allocation dynamics. This attainable set determines the magnitude and bandwidth constraints of the FES-induced torque. 
The magnitude constraints of FES-induced torque in this simulation are derived from real experimental data.
Moreover, to better highlight the impact of dynamic discrepancy and illustrate how the dynamic allocator accounts for it, the bandwidths of the biceps and triceps muscle groups are set to \(0.908\,\text{Hz}\) and \(3.976\,\text{Hz}\), respectively.

\begin{figure}[!ht]
    \centering
    \vspace*{0mm}
        \begin{subfigure}[b]{0.49\textwidth}
            \centering
            \includegraphics[width=\textwidth,clip,trim={5.5cm 13.1cm 6.8cm 12.0cm}]{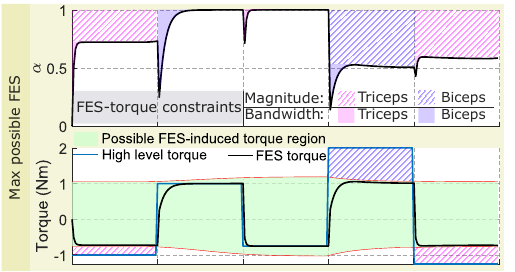}
            \put(-237,120){\textbf{(a)}} 
            \phantomcaption
            \label{fig:DynamicAllocationSimulation_m}
        \end{subfigure}
        \begin{subfigure}[b]{0.49\textwidth}
            \centering
            \includegraphics[width=\textwidth,clip,trim={5.5cm 8.0cm 6.8cm 16.6cm}]{Fig/examplar_2_7.pdf}
            \put(-237,132){\textbf{(b)}} 
            \phantomcaption
            \label{fig:DynamicAllocationSimulation_b}
        \end{subfigure}
    \vspace*{-2mm}
    \caption{\textbf{Simulation results of the dynamic allocation.} The allocation considers the attainable sets and constraints (maximum torque magnitude and bandwidth limits) of the actuators when distributing the control effort.
    Two scenarios are presented, both aiming to maximize FES usage (\(80\%\) and \(100\%\)) within its attainable set. \textbf{(a)} Maximum FES usage within its attainable set. \textbf{(b)} Up to \(80\%\) FES assistance within its attainable set.}
    \label{fig:DynamicAllocationSimulation}
     \vspace*{-0mm}
\end{figure}

\textbf{Results:}
The simulation illustrates two control scenarios. In the first, the objective is to maximize the use of FES-induced torque (Fig.~\ref{fig:DynamicAllocationSimulation_m}). In the second scenario, a minimum contribution of \(20\%\) exoskeleton contribution is considered, effectively limiting the maximum FES share to \(80\%\) (Fig.~\ref{fig:DynamicAllocationSimulation_b}). In  both cases, the dynamic allocation, \eqref{eq:allocator_structure_a} and \eqref{eq:zeta_eq_2}, balances contributions from the exoskeleton and FES, ensuring compliance with FES dynamics and constraints.
%
%
Although abrupt changes in the high-level desired torque are generally avoided in practice, they were introduced in this simulation to test the allocator’s robustness under demanding conditions. The results show that, even in such cases, the FES torque varies smoothly and remains within the attainable set, satisfying the bandwidth constraints during transitions.
The faster resource allocation to the triceps compared to the biceps is attributed to differences in the maximum attainable bandwidths of the low-level FES control considered in the FES attainable set (\(0.908\)~Hz for biceps and \(3.976\)~Hz for triceps).

    \vspace*{-1mm}
\noindent

\subsubsection{Input Redundancy and Dynamic Allocation within Muscle Groups} \label{sec:Input Redundancy and Dynamic Allocation within Muscle Groups}
So far, we have addressed dynamic allocation between FES and the exoskeleton, considering the FES flexor and extensor each as a single actuator. We now extend the dynamic allocation to also account for redundancy \textit{within} muscle groups.
\begin{assumption} \label{assump:assumption_1}
Each individually stimulated muscle is modeled as a separate actuator, and the contributions of ligaments and articular surface contact forces are negligible~\cite{crowninshield1981physiologically}.
\end{assumption}

Under Assumption~\ref{assump:assumption_1}, the flexor and extensor FES torques can be expressed as
\begin{equation} \label{eq:torque_cc_2}
\begin{split}
&\tau^{F_f} = \boldsymbol{1}_{n_f}^\top  \boldsymbol{\tau}^{\boldsymbol{m_f}},\\
&\tau^{F_e} = \boldsymbol{1}_{n_e}^\top  \boldsymbol{\tau}^{\boldsymbol{m_e}},
\end{split}
\end{equation}
\noindent where \(\mathbf{1}_{\rho} \in \mathbb{R}^{\rho}\) denotes the column vector of ones of appropriate dimension \(\rho\). \(\boldsymbol{\tau}^{\boldsymbol{m_f}} \in \mathbb{R}^{n_{mf}}\) and \(\boldsymbol{\tau}^{\boldsymbol{m_e}} \in \mathbb{R}^{n_{me}}\) are the vectors of torque contributions from all flexor and extensor muscle groups in response to FES contributing on joint \(j\), respectively. Explicitly,  
\begin{equation} \label{eq:torque_muscle_groups}
\begin{split}
    &\boldsymbol{\tau}^{\boldsymbol{m_f}} = [{\tau}^{{m}_{f_1}}, {\tau}^{{m}_{f_2}}, ..., {\tau}^{{m}_{f_{n_{f}}}}]^\top ,\\
    &\boldsymbol{\tau}^{\boldsymbol{m_e}} = [{\tau}^{{m}_{e_1}}, {\tau}^{{m}_{e_2}}, ..., {\tau}^{{m}_{e_{n_{e}}}}]^\top ,
\end{split}
\end{equation}
where \(n_{f}\) and \(n_{e}\) denote the number of flexor and extensor muscle groups inducing torque at joint \(j\), respectively.
Each muscle generates only positive tension, but due to opposite moment arms, flexor torques appear with a positive sign and extensor torques with a negative sign in the chosen joint coordinate system.


The proposed dynamic allocation method is capable of exploiting such redundancy. To reflect this, we redefine the torque vector in~\eqref{eq:torque_vec} to represent the torque contributions from each muscle group separately. Given~\eqref{eq:torque_muscle_groups}, we express the extended torque vector as
\begin{equation} \label{eq:torque_vec_extended}
\begin{split} 
    \boldsymbol{\tau}^{\scriptscriptstyle +} = \begin{bmatrix}
    {\tau}^{{m}_{f_1}}, \ldots, {\tau}^{{m}_{f_{n_{f}}}}, 
    {\tau}^{{m}_{e_1}}, \ldots, {\tau}^{{m}_{e_{n_{e}}}}, 
    {\tau}^{E}
    \end{bmatrix}^\top,
\end{split}
\end{equation}
where \(\boldsymbol{\tau}^{\scriptscriptstyle +}\) includes the torques from all flexor muscles, extensor muscles, and the exoskeleton actuator at a given joint \(j\). The superscript “\( {\scriptscriptstyle +} \)” is used here to denote the extended structure that distinguishes this formulation from earlier definitions (Section~\ref{section: Input redundancy in hybrid exoskeleton} and~\ref{section: Dynamic allocation in hybrid exoskeleton}).

Similar to \eqref{eq:allocator_structure}, the governing equations of the hybrid FES-exoskeleton dynamic allocator can now be expressed as
\begin{subequations} \label{eq:allocator_structure_extended}
\begin{align}
    \boldsymbol{\tau}^{\scriptscriptstyle +} &= \bar{\boldsymbol{\tau}}^{\scriptscriptstyle +} + \boldsymbol{g}^{*{\scriptscriptstyle +}}_{\perp} S^{\scriptscriptstyle +} \boldsymbol{\zeta}^{\scriptscriptstyle +}, \label{eq:allocator_structure_extended_a} \\
    \dot{\boldsymbol{\zeta}}^{\scriptscriptstyle +} &= \boldsymbol{\phi}^{\scriptscriptstyle +}(\boldsymbol{\zeta}^{\scriptscriptstyle +}, \bar{\boldsymbol{\tau}}^{\scriptscriptstyle +}, \boldsymbol{\eta}^{\scriptscriptstyle +}), \label{eq:allocator_structure_extended_b} \\
    \text{s.t.} \quad &
    \begin{cases}
        {\tau}^{{m}_{f_i}} \in \mathbb{A}_{{m}_{f_i}}, \\
        {\tau}^{{m}_{e_i}} \in \mathbb{A}_{{m}_{e_i}}, \\
        {\tau}^E \in \mathbb{A}_E,
    \end{cases} \nonumber
\end{align}
\end{subequations}

\noindent where \(\dot{\boldsymbol{\zeta}^{\scriptscriptstyle +}} = \boldsymbol{\phi}^{\scriptscriptstyle +}(.)\) is input-to-state stable with respect to \(\bar{\boldsymbol{\tau}}^{\scriptscriptstyle +}(t)\) and
\[
\boldsymbol{g}^{*{\scriptscriptstyle +}}_{\perp} = \begin{bmatrix}
\mathbf{I}_{n_m} \\
-\mathbf{1}_{n_m}^\top
\end{bmatrix} \in \mathbb{R}^{(n_m+1) \times n_m},
\]
\( n_m = n_f + n_e \) denotes the total number of muscle groups and \( S^{\scriptscriptstyle +} =S \begin{bmatrix}
\mathbf{1}_{n_f}^\top & \mathbf{0}_{n_e}^\top \\
\mathbf{0}_{n_f}^\top & \mathbf{1}_{n_e}^\top
\end{bmatrix}\)
, and \( \boldsymbol{\zeta}^{\scriptscriptstyle +} = [\zeta^{\scriptscriptstyle +}_1, \zeta^{\scriptscriptstyle +}_2, \ldots, \zeta^{\scriptscriptstyle +}_{n_m}]^\top \in \mathbb{R}^{n_m} \) is the dynamic input allocator, with \( l^{\scriptscriptstyle +} = \mathrm{rank}(\boldsymbol{g}^{*{\scriptscriptstyle +}}_{\perp}) = n_m \).
Moreover, \(\mathbf{I}_{\rho} \in \mathbb{R}^{\rho \times \rho}\) and \(\mathbf{0}_{\rho} \in \mathbb{R}^{\rho}\) denote the identity matrix and the column vector of zeros of appropriate dimension \(\rho\), respectively. 

Similar to \eqref{eq:zeta_eq_2} and Theorem~\ref{theorem2}, it is straightforward to show that the allocation dynamics
\begin{equation} \label{eq:zeta_eq_extended}
\begin{split}
    \dot{\boldsymbol{\zeta}}^{\scriptscriptstyle +} = -K^{\scriptscriptstyle +} S^{\scriptscriptstyle +} {\boldsymbol{g}^{*{\scriptscriptstyle +}}_{\perp}}^\top W^{\scriptscriptstyle +} \boldsymbol{\tau}^{\scriptscriptstyle +},
\end{split}
\end{equation}
\noindent is a valid candidate for the dynamic allocation structure in~\eqref{eq:allocator_structure_extended}. Here, \( K^{\scriptscriptstyle +} = \mathrm{diag}([k_1^{\scriptscriptstyle +}, \ldots, k_{n_m}^{\scriptscriptstyle +}]) \in \mathbb{R}^{n_m \times n_m} \) and \( W^{\scriptscriptstyle +} = \mathrm{diag}([w_1^{\scriptscriptstyle +}, \ldots, w_{n_m+1}^{\scriptscriptstyle +}]) \in \mathbb{R}^{(n_m+1) \times (n_m+1)} \) are positive diagonal matrices that determine the convergence speed and distribution ratio of the dynamic allocation control.

\begin{remark}[Limitation of Dynamic Allocation \textit{within} Muscle Groups]  \label{Limitation of Dynamic Allocation within Muscle Groups}
Although dynamic allocation provides a theoretical solution for addressing input redundancy \textit{within} muscle groups~\ref{sec:Input Redundancy and Dynamic Allocation within Muscle Groups}, its practical implementation would require isolated observation of individual muscle responses, which remains impractical with current experimental techniques. Nevertheless, this formulation, \eqref{eq:allocator_structure_extended} and \eqref{eq:zeta_eq_extended}, provides valuable insight into redundancy \textit{within} muscle groups and a principled framework for resolving redundancy in musculoskeletal models (e.g., in OpenSim and MuJoCo).
\remarkend 
\end{remark}

\subsection{FES Control}\label{sec:fes_control}
\revisionStatus{1}
Distribution of control effort between exoskeleton and FES based on their constraints, along with precise low-level control of the nonlinear~\cite{kirsch2017nonlinear, chang2020model}, time-varying~\cite{alibeji2017adaptive,sena2022gap} FES dynamics, which vary across individuals~\cite{kavianirad2023model, kavianirad2024toward}, necessitates the development of FES-torque model~\cite{chen2014fes, durfee1989methods, kavianirad2023model}. This model serves two key functions: first, it determines the attainable FES sets \(\mathbb{A}_F\) used in dynamic allocation of high-level control and, second, the FES model \(\Sigma_{F}\) used in low-level control to ensure accurate FES-induced torque generation.



\subsubsection{FES-Torque Model}\label{section:FES-torque model}
Several FES models describing the neuromuscular response to artificial stimulation are introduced in the literature~\cite{chen2014fes, cousin2018admittance, alibeji2018control}. In this study, one of the commonly used FES models incorporating activation and contraction dynamics is used. The FES-induced muscle torque \(\tau^{F}\) in this model is the product of activation and contraction dynamics~\cite{chen2014fes, romero2019design, kavianirad2023model}. The contraction dynamics defines the maximum torque that can be induced at a certain joint angle \(\theta\), denoted by \(\tau^{F^*}(\theta)\), while the activation \(a_{\psi}\) determines the extent to which FES recruits a motor unit
\begin{equation}
    \tau^{F} ={a_{\psi}}\tau^{F^*}~.\label{eq:tauFES}
\end{equation}
The activation dynamics itself consists of static nonlinear recruitment, linear dynamics (calcium released dynamics~\cite{chen2014fes}) representing the excitation of artificially stimulated muscles, delay~\cite{allen2023electromechanical}, and time-varying fatigue~\cite{sena2022gap, chen2014fes, bao2020tube}
\begin{subequations} \label{eq:activation}
\begin{align}
    & a_{\psi}=\psi a(t-t_d), \label{eq:fatigue_delay}\\
    & \dot{\boldsymbol{a}}=A{\boldsymbol{a}}+Ba_r~, \label{eq:calcium}\\
    & a_r=r(\upsilon,\theta)~, \label{eq:recruitment}
\end{align}
\end{subequations}
\noindent where \({\psi}\) and \({t_d}\) represent the FES fatigue coefficient and delay effect into the model, and \({A} \in \mathbb{R}^{2 \times 2}\) and \({B} \in \mathbb{R}^{2 \times 1}\) define non-fatigue activation dynamics with \({\boldsymbol{a}}=[{a},\dot{a}]^\top \). Moreover, \(\upsilon\), \(a_r\), and \(r(.\)) are FES control input (stimulation intensity), recruitment characteristic, and the nonlinear recruitment curve, respectively~\cite{chen2014fes, dunkelberger2022shared}.

\subsubsection{FES-Torque Model Identification}\label{section:FES-torque model identification}

The static maps \( \tau^{F^*}(.) \) and \( r(.) \), as well as the delay \( t_d \) and the dynamic map parameters \( A \) and \( B \), are unknown, requiring individual identification. Hammerstein-Wiener system identification~\cite{wills2013identification}, therefore, is used to identify the FES-torque model~\cite{hunt1998investigation, schauer2005online, winter2009biomechanics, kavianirad2023model} based on training data. The nonlinear muscle recruitment curve is modeled by the cubic spline~\cite{dempsey2004identification, kavianirad2023model, alibeji2018muscle, tu2017upper, sartori2012modeling, le2010identification, riener1996biomechanical}, and the linear activation dynamics~\eqref{eq:calcium} is then learned using the linear regression method~\cite{hunt1998investigation, winter2009biomechanics, kavianirad2023model}. Moreover, attainable sets of FES \(\mathbb{A}_F\) are derived from the learned map.

Training data for the FES-induced torque model consists of FES torque at various FES intensities at different elbow angles 
\begin{equation} \label{eq:training_data_set}
    \left\{ \left( [\upsilon^{(\gamma)}, \; \theta^{(\gamma)}]^\top, \; {\tau^{F}}^{(\gamma)}\right) \right\}_{\gamma=0}^{\Gamma}
\end{equation}
where the superscript \( (\gamma) \) indicates the value of each variable at the \( \gamma \)-th sample, with \( \gamma \in \{0, 1, \ldots, \Gamma\} \), and \( \Gamma \) being the total number of training samples.


In identification, the cumulative delay comprising electromechanical delay~\cite{chen2014fes, allen2023electromechanical}, actuator/stimulator delay, and communication delay is estimated. The stimulator and communication delays depend on the FES and control setup and are estimated \(16 ms \) for the system in this study. 
Moreover, fatigue modeling and identification are not investigated in this work; nevertheless, the provided framework allows for incorporating this effect (see~\cite{chen2014fes, bao2020tube} for fatigue modeling). Fatigue limits the FES-induced torque and, therefore, the attainable set of FES \(\mathbb{A}_F\), this can be taken into account by updating FES magnitude saturation, \(W\) in \eqref{eq:zeta_eq_2}; further details on incorporating saturation into \( W \) are provided in~\cite{zaccarian2009dynamic}.

\subsubsection{FES Torque Control}
To evaluate the performance of the hybrid system under the proposed cooperative control method and minimize the influence of other factors, such as the FES-induced torque observer required in low-level closed-loop FES torque control, a feedforward low-level FES control based on the learned FES model (described in Sections~\ref{section:FES-torque model} and~\ref{section:FES-torque model identification}) is employed.
The learned static map (\(\tau^{F^*}(\theta)\), \(r(\upsilon,\theta)\)), the dynamic map parameters (\(A\), \(B\)), the desired FES torque~\(\tau^{F}\) determined by the cooperative control, and the measured joint angle~\(\theta\) are the inputs to the low-level FES controller.

Given \eqref{eq:tauFES}, the desired activation \(a_{\psi}\) is determined from the desired FES torque and \(\tau^{F^*}(\theta)\) at the measured angle. Based on the activation dynamics defined in~\eqref{eq:calcium} and using a feedforward control approach, the desired recruitment characteristic \(a_r\) is determined based on \({a}_{\psi}\). To normalize the FES intensity from recruitment characteristic $a_r$, the solution of the static map \eqref{eq:recruitment} is required. This is achieved by minimizing the cost function $J=(r(\upsilon,\theta)-a_r)^2$, 
\begin{equation} \label{eq:des_torque}
\begin{split}
&{\upsilon}^{\star}=\arg\min_{\upsilon} J(\upsilon,\theta,a_r) \\
&\;\;\;\;\text{s.t.}\;\;\;\;0 \leq \upsilon \leq \upsilon_{max}~,
\end{split}
\end{equation}
\noindent where \(\upsilon_{\text{max}}\) denotes the maximum-but-comfortable FES intensity threshold.

\subsection{Exoskeleton Control}\label{sec:exo_control}
\revisionStatus{1}
The adopted exoskeleton is torque-controlled. The following torque command is sent to the low-level exoskeleton torque controller
\begin{equation} \label{eq:cc}
    {\tau^{E'}} = \tau^{E} + {\tau}^{\text{g}}~,
\end{equation}
\noindent where \(\tau^{E}\) denotes the assistive exoskeleton torque determined by the cooperative control and \({\tau}^{\text{g}}\) represents the gravity torque, which accounts for the weight of both the robot and the participant's arm. Feedforward gravity compensation allows motion not to be biased in the direction of gravity~\cite{kavianirad2023model, de2006collision}, thereby guaranteeing safer operation of the system~\cite{heinzmann2003quantitative}. The low-level exoskeleton control used the pole-placement method, based on the dynamic model of the actuators, to ensure torque control precision and stability~\cite{pan2022nesm}. 

\section{Experimental Evaluation} \label{sec:Experimental Evaluation}

To evaluate the proposed dynamic torque allocation, we conducted a performance assessment with the hybrid FES-exoskeleton introduced in~\ref{sec:Hybrid FES-Sxoskeleton Setup}.
The proposed cooperative control allows for online adaptation of the control distribution \(\alpha\) based on learned FES-induced torque constraints~(\ref{section: Model_Identification}). In this demonstration, we designed the allocation so that it prioritizes the more constrained assistive device, namely FES, over the exoskeleton, reflecting both ideal clinical practice and a more challenging control condition. Its performance was then compared against a constant allocation method. 

%

In both dynamic and constant allocation, the tracking task was to follow the elbow joint reference trajectory. The reference trajectory was generated by introducing a constant phase of \(2\) seconds at the local extrema of the following trajectory, resulting in a combination of constant and time-varying reference trajectories
\begin{equation}
\begin{split}
    & {\theta_{d_0}} = \theta_0 + \theta_A \prod_{i=1}^{3} \sin(2\pi f_i (t-t_0)), 
    \label{eq:theta_d}
\end{split}
\end{equation} 
\noindent where \(t\) denotes time, and \(t_0\) is a time offset. The parameters are set as \(t_0=2.5\), \(\theta_0 = 52.6^\circ\), \(\theta_A = 37.6^\circ\), and frequency components \(f_1 = 0.050\)~Hz, \(f_2 = 0.068\)~Hz, and \(f_3 = 0.093\)~Hz. The time \(t\) is defined over the interval \(0 \leq t \leq 30\) seconds. The desired trajectory is depicted in Fig.~\ref{fig:exemplar_1_c} (top). 
Moreover, for a fair comparison, the constant \(\alpha\) used in the constant allocation was set to the average \(\alpha\) from the dynamic allocation trial. 

%
The Research Protocol of this study was approved by the Ethics Committee of the Scuola Superiore Sant'Anna (approval n. 18/2024), following the principles stated in the Declaration of Helsinki.

\subsection{Hybrid FES-Exoskeleton Setup}\label{sec:Hybrid FES-Sxoskeleton Setup} 

\subsubsection{Functional Electrical Stimulation}
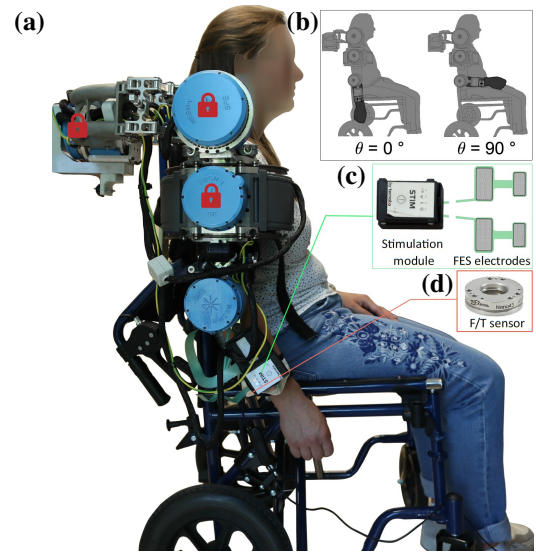
\begin{figure}[ht]
    \centering
        \includegraphics[width=0.40\textwidth,angle=90,trim={0.2cm 0.2cm 0.1cm 0.1cm},clip]{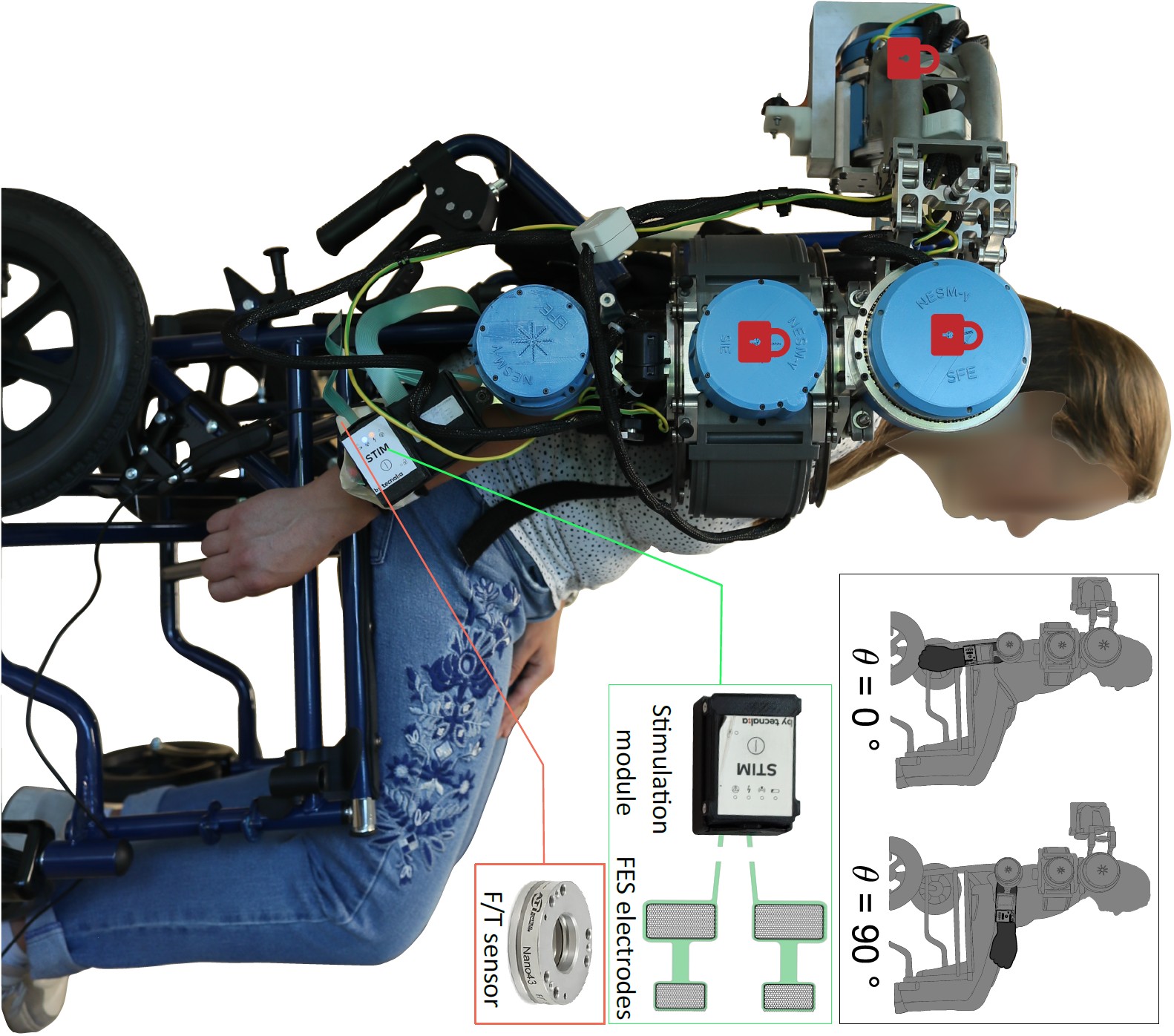}
    \vspace*{-0mm}
    \put(-200,197){\textbf{(a)}} 
    \put(-97,197){\textbf{(b)}} 
    \put(-78,139){\textbf{(c)}} 
    \put(-46,98){\textbf{(d)}} 
    \caption{\textbf{Illustration of the hybrid FES-exoskeleton.} \textbf{(a)} 
    Only the elbow joint was used to follow a trajectory.
    \textbf{(b)} The degree is defined so that \(0^\circ\) and \(90^\circ\) correspond to the elbow angle at which the forearm points downwards and is parallel to the ground, respectively. \textbf{(c)} The FES stimulation is applied to the elbow extensor (triceps brachii) and flexor (brachialis). \textbf{(d)} The F/T sensor is mounted at the interface between the exoskeleton and the user’s forearm, where it measures interaction forces and torques.
    }
    \label{fig:exo_fes}
\end{figure}

A research-grade FES device (Tecnalia Research \(\&\) Innovation, Spain), interfaced with multi-array electrodes (Fig.~\ref{fig:exo_fes}), provides real-time control over stimulation parameters, including pulse frequency, width, and amplitude. In this study, FES delivers a biphasic electrical pulse at a fixed frequency of \(25\)~Hz and a fixed pulse width of \(300\)~\(\mu s\). The pulse amplitude is used as the control variable and is adjustable with a command resolution of \(100\)~\(\mu A\).

\subsubsection{Exoskeleton}
NESM-\(\gamma\)~\cite{pan2022nesm, penna2024muscle, pan2023self}, depicted in Fig.~\ref{fig:exo_fes}, is a powered shoulder-elbow exoskeleton using series-elastic actuators to assist upper-limb movement. The exoskeleton features a passive kinematic chain to maintain alignment of the robot with the human shoulder joint, and four active revolute joints driven by series elastic actuators: three at the shoulder, for adduction/abduction, flexion/extension, and intra/extra rotation, and one at the elbow for flexion/extension. The control unit of the exoskeleton comprises a real-time controller, sbRIO-9651 (National Instruments—NI, Austin, TX, USA), endowed with a Xilinx Zynq-7020 System on Chip, combining a real-time (RT) processor and a Field Programmable Gate Array (FPGA). The high-level control runs at \(100\)~Hz on the RT unit, while the low-level control runs at \(1\)~kHz on the FPGA unit~\cite{pan2022nesm}. A LabVIEW Graphical User Interface (GUI) allows online interaction with the RT level, to set operator's commands and control parameters (e.g., the arm gravity compensation tuning). A Nano-43 force/torque (F/T) sensor (ATI Industrial Automation, USA) is integrated by means of an adjustable mechanical interface connecting the exoskeleton to the user's forearm to measure interaction forces and torques. Force and torque measurements are mapped into elbow interaction torque, according to the subject-specific distance between the sensor and the elbow joint. To focus on cooperative control and redundant actuation, in this study, the passive joints are mechanically locked and the shoulder joints are controlled in a configuration that holds the upper arm in a vertical position, while the hybrid FES-exoskeleton assists at the elbow level, as shown in Fig.~\ref{fig:exo_fes}. 

\subsection{Model Identification}\label{section: Model_Identification}
FES model is identified based on~\ref{section:FES-torque model identification}. 
Training data set~\eqref{eq:training_data_set} for the FES-induced torque model consists of providing five FES current intensities~(equally distributed between minimum and maximum intensities and randomized in order) at six elbow angles~(\(15^\circ\), \(30^\circ\), \(45^\circ\), \(60^\circ\), \(75^\circ\), and \(90^\circ\)) for both elbow flexor (biceps brachii) and extensor (triceps brachii). 
The exoskeleton automatically guides the elbow to the desired joint angle. This movement is performed slowly with rest periods between transitions to avoid increased joint stiffness. For each intensity and angle, torque measurements are obtained via the embedded F/T sensor. In total, the training dataset comprises \(60\) stimulations for each muscle group, with each stimulation lasting \(5\) seconds and interleaved with a \(5\)-second resting period.


\subsection{Procedure}

At the start, the FES electrodes and the exoskeleton are fitted to the participant, and a calibration of the hybrid system is performed before the tracking task. Calibration is used to learn the FES model and the attainable FES set, as well as to adjust the exoskeleton’s gravity compensation. The calibration and tracking task are conducted while the participant is passively sitting in a chair. FES calibration begins with manually adjusting the stimulation intensity to determine the maximum comfortable threshold and the minimum threshold that elicits contractions in the targeted muscles (elbow flexor/extensor). Next, gravity compensation is tuned to compensate for the weight of the exoskeleton components (robot's structure, F/T sensor, and interfaces) as well as the person’s arm. 
This prevents motion from being biased by gravity, which would otherwise result in predominantly flexor activation and limit the evaluation. Then, the FES model is generated before the tracking task.

\section{Results} \label{sec:Results}
\begin{figure*}[!ht]
    \centering
    \vspace*{0mm}

    \begin{subfigure}[b]{0.26\textwidth}
        \centering
        \begin{subfigure}[b]{\textwidth}
            \centering
            \includegraphics[width=\textwidth,clip,trim={0.4cm 24.2cm 15.3cm 1cm}]{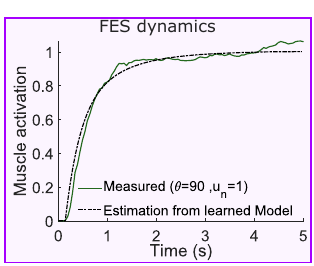}
            \put(-130,97){\textbf{(a)}} 
            \phantomcaption
            \label{fig:exemplar_1_a}
        \end{subfigure}
        \begin{subfigure}[b]{\textwidth}
            \centering
            \includegraphics[width=\textwidth,clip,trim={0.4cm 20.0cm 15.3cm 5.5cm}]{Fig/examplar_2_4_6_tcst.pdf}
            \put(-130,94){\textbf{(b)}} 
            \phantomcaption
            \label{fig:exemplar_1_b}
        \end{subfigure}
    \end{subfigure}
    \begin{subfigure}[b]{0.41\textwidth}
        \centering
        \includegraphics[width=\textwidth,clip,trim={5.8cm 20.0cm 7.0cm 1cm}]{Fig/examplar_2_4_6_tcst.pdf}
        \put(-213,206){\textbf{(c)}} 
        \phantomcaption
        \label{fig:exemplar_1_c}
    \end{subfigure}
    \begin{subfigure}[b]{0.30\textwidth}
        \centering
        \begin{subfigure}[b]{\textwidth}
            \centering
            \includegraphics[width=\textwidth,clip,trim={14.02cm 25.6cm 1.0cm 1cm}]{Fig/examplar_2_4_6_tcst.pdf}
            \put(-160,58){\textbf{(d)}} 
            \phantomcaption
            \label{fig:exemplar_1_d}
        \end{subfigure}
        \begin{subfigure}[b]{\textwidth}
            \centering
            \includegraphics[width=\textwidth,clip,trim={14.02cm 20.5cm 1.0cm 4.24cm}]{Fig/examplar_2_4_6_tcst.pdf}
            \phantomcaption
            \label{fig:exemplar_1_e}
        \end{subfigure}
        \put(-160,142){\textbf{(e)}} 
        
    \end{subfigure}

    \vspace*{-5mm}
    \caption{\textbf{Experimental results of the hybrid FES-exoskeleton system under dynamic allocation.} \textbf{(a)} FES activation dynamics indicating actuator bandwidth constraints. \textbf{(b)} FES static torque map representing actuator magnitude saturation. Training data and learned static map for both muscle groups are shown. The static map illustrates both the biceps map (positive \(\upsilon\)) and triceps map (negative \(\upsilon\)). \textbf{(c)} Exemplary result demonstrating trajectory tracking, cooperative gain, high-level desired torque \(\tau^N\), and desired FES torque \(\tau^F\). The green shaded area represents the attainable set of FES torque \(\mathbb{A}_F\), derived from the FES torque map in \textbf{(b)} for this specific tracking task. \textbf{(d)} Cooperative gain distribution. \textbf{(e)} Compliance with actuator bandwidth constraints (FES dynamic response in \textbf{(a)}) and magnitude saturation (FES torque map in \textbf{(b)}).}
    \label{fig:exemplar_1}
\end{figure*}

\revisionStatus{1}
Fig.~\ref{fig:exemplar_1} illustrates an exemplar result of the dynamic allocation in the hybrid system. The attainable actuator set used in the dynamic allocation scheme is based on the muscle activation dynamics (capturing FES bandwidth constraints) and the static FES torque map (describing actuator magnitude saturation) shown in Fig.~\ref{fig:exemplar_1_a} and Fig.~\ref{fig:exemplar_1_b}, respectively. 


Fig.~\ref{fig:exemplar_1_c} (top) demonstrates the joint trajectory tracking result, where the measured joint angle \(\theta\) closely follows the reference \(\theta_d\), with a Root Mean Square Error (RMSE) of \(3.76^\circ\). Fig.~\ref{fig:exemplar_1_c} (middle) depicts the cooperative gain \(\alpha\) modulated adaptively based on user preferences (prioritizing FES over exoskeleton) and attainable sets of actuators (Fig.~\ref{fig:exemplar_1_a} and Fig.~\ref{fig:exemplar_1_b}). 
We designed the control scenario such that FES would be the preferred assistive technology, and the distribution of the cooperative gain, shown in Fig.~\ref{fig:exemplar_1_d}, supports effective integration of this consideration within dynamic allocation: \(\alpha\) distribution is skewed toward higher values, with \( P(\alpha \geq 0.95) \approx 64\% \) and average \(\alpha \approx 0.90\), which indicates prioritization of the FES over exoskeleton. Fig.~\ref{fig:exemplar_1_c} (bottom) shows the desired assistive torque, the desired FES torque, and the attainable FES-induced torque set, derived from the static map in Fig.~\ref{fig:exemplar_1_b}. The desired FES torque lies within the green shaded region, indicating compliance with actuator magnitude saturation constraints throughout the whole trial. Fig.~\ref{fig:exemplar_1_e} presents zoomed-in segments that confirm compliance with both magnitude and bandwidth constraints. These results highlight the ability of the proposed dynamic allocation framework to prioritize one actuator over another while respecting actuator constraints and ensuring accurate trajectory tracking.


\begin{figure}[!ht]
     \centering
     \vspace*{0mm}
\includegraphics[width=0.49\textwidth,clip,trim={0.02cm 19.7cm 11.00cm 0.00cm}]{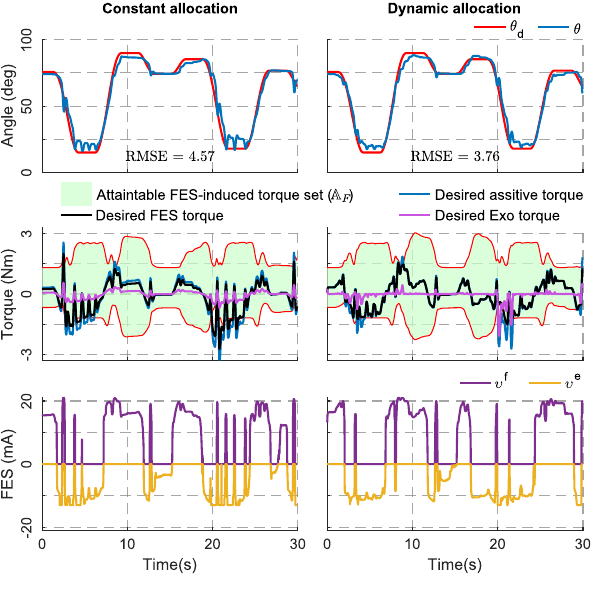}
      \vspace*{-5mm}
         \caption{\textbf{Experimental results of the hybrid FES-exoskeleton system (comparison of constant and dynamic allocation).} This figure shows the desired and measured elbow angle~(\(\theta_{d}\) and \(\theta\)), attainable FES-induced torque set \(\mathbb{A}_F\), desired assistive torque, desired FES torque, desired exoskeleton torque, and flexor/extensor FES control profiles (\(\upsilon^{f}\) and \(\upsilon^{e}\)). Dynamic allocation demonstrates better tracking performance compared to the constant allocation while ensuring compliance with the FES attainable set.}
     \vspace*{-3mm}
     \label{fig:exemplar_all_1}  
\end{figure}

Fig.~\ref{fig:exemplar_all_1} compares the performance of the hybrid system under constant and dynamic allocation.
As discussed in~\ref{sec:Experimental Evaluation}, \(\alpha\) used in the constant allocation is equal to the average \(\alpha\) from the dynamic allocation trial.
The dynamic allocation achieves higher tracking accuracy (RMSE of \(3.76^\circ\) versus \(4.57^\circ\) for constant allocation) while respecting FES constraints. A closer inspection of the desired FES torque in the constant allocation condition reveals instances where the desired FES torques violate actuation feasibility. In contrast, the dynamic allocation consistently satisfies these constraints, ensuring that the desired FES torque remains within its attainable set.



\section{Conclusion} \label{sec:Conclusion}
\revisionStatus{0}
The present study introduces and evaluates a novel adaptive cooperative control architecture based on dynamic allocation to solve actuator redundancy in a hybrid FES-exoskeleton. Dynamic allocation allows the efficient integration of FES and exoskeleton through real-time distribution of control effort between assistive technologies with different dynamics and constraints based on user preferences, control performance, and secondary objectives such as rehabilitation uptake. Simulation results and experimental validation confirm the effectiveness of the proposed method in addressing actuator redundancy. Although the adaptive cooperative control approach is not limited to any specific prioritization, it enables the system to prioritize FES—the more constrained yet preferred assistive technology—as the primary source of assistance, with the exoskeleton acting as a complementary assistive device. Experimental results support this FES-prioritizing strategy, demonstrating that, on average, up to \(90\%\) of the needed assistive torque can be provided by FES while still ensuring user safety and comfort and adhering to the capabilities and constraints of the assistive devices.

This study focused on a single joint, but the proposed approach is straightforward to generalize to other joints. Furthermore, co-contraction and the null-space basis defined here correspond to a one-degree-of-freedom joint motion (e.g., elbow); for more complex joint motions, these formulations would need to be adapted accordingly. 
%
%
Moreover, while the method allows for adaptation to muscle fatigue as a time-varying magnitude saturation, fatigue was not considered in the experimental evaluation. Similarly, sensor and model noise were not explicitly incorporated in allocation. These factors represent directions for future extensions of the dynamic allocation in FES–exoskeleton systems.

\appendices

\section{Proof of Theorem \ref{theorem:invisibility}: Invisibility of Redistribution Term in Dynamic Allocation}\label{sec:Apendix_A}
To show that the redistribution term \(\boldsymbol{g}^*_{\perp} S \boldsymbol{\zeta}\) is invisible to the hybrid high-level controller and the system output, we need to verify that distributed torque components are bounded in the first place and the system states and outputs with input \(\boldsymbol{\tau} = \bar{\boldsymbol{\tau}} + \boldsymbol{g}^*_{\perp} S \boldsymbol{\zeta}\) are identical to those obtained under the nominal input \(\bar{\boldsymbol{\tau}}\). 

From~\eqref{eq:allocator_structure_a}, the control input norm satisfies
\[
\|\boldsymbol{\tau}(t)\| \leq \|\bar{\boldsymbol{\tau}}(t)\| + \|\boldsymbol{g}_{\perp}^* S\| \|\boldsymbol{\zeta}(t)\|,
\]
under Assumption~\ref{assump:internal_stability}, \( \|\boldsymbol{\zeta}(t)\| \) is bounded by~\eqref{eq:ISS_def}, therefore
\[
\|\boldsymbol{\tau}(t)\| \leq \|\bar{\boldsymbol{\tau}}(t)\| + \|\boldsymbol{g}_{\perp}^* S\| \left( \delta(\|\boldsymbol{\zeta}(0)\|, t) + \mu\left( \|\bar{\boldsymbol{\tau}}\|_{[0,t]} \right) \right).
\]
by assumption the nominal control input \( \bar{\boldsymbol{\tau}}(t) \) is uniformly bounded for all \( t \geq 0 \), i.e., there exists a constant \( \bar{\tau}_{\max} > 0 \) such that \( \|\bar{\boldsymbol{\tau}}(t)\| \leq \bar{\tau}_{\max} \), it follows that
\[
\|\boldsymbol{\tau}(t)\| \leq \bar{{\tau}}^{\max} + \|\boldsymbol{g}_{\perp}^* S\| \left( \delta(\|\boldsymbol{\zeta}(0)\|, t) + \mu(\bar{{\tau}}^{\max}) \right).
\]
which implies that \(\boldsymbol{\tau}(t)\) is uniformly bounded for all \(t \geq 0\). Therefore, all actuator torques remain bounded. 

Finally, since \(\boldsymbol{g}^*_{\perp}\) lies in the null space of the augmented control matrix of the hybrid system, the redistribution term \(\boldsymbol{g}^*_{\perp} S \boldsymbol{\zeta}\) does not affect the system dynamics. Therefore, under the given assumption, the redistribution input \(\boldsymbol{g}^*_{\perp} S \boldsymbol{\zeta}\) is invisible to the system's high-level controller and system output, which confirms~\eqref{eq:IR_mathmatically_2}.

\section{Internal Stability of Switching Allocation Dynamics}\label{sec:Apendix_B}
Based on the definition of \(S\) or \(\sigma_i\), the system defined in~\eqref{eq:zeta_eq_2} or \eqref{eq:zeta_eq3} constitutes a time-dependent switching system~\cite{liberzon2003switching}, since the sign of \(\bar{{\tau}}^N(t)\) determines the dynamics of the allocation. Note that, given~\eqref{eq:allocator_structure_a} and~\eqref{eq:g_orth}, the sign of \(\bar{\tau}^N(t)\) equals the sign of \({\tau}^N(t)\), so the two can be used interchangeably.
Given~\eqref{eq:zeta_eq_2}, we can write

\begin{equation} \label{eq:zeta1_dynamics}
    \dot{\zeta}_i = -b_{i\sigma} \left( \zeta_i - c_i\bar{\tau}^N \right), \quad i = 1,2
\end{equation}
where
\begin{equation} \label{eq:bsigma_and_c}
\begin{split}
    b_{i\sigma}(t) &=
    \begin{cases}
        k_i(w_i(t) + w_3(t)), & (-1)^i \bar{\tau}^N \prec 0, \\
        0, &  \text{otherwise},
    \end{cases} \\
    c_i(t) &= \frac{w_3(t)}{w_i(t) + w_3(t)} - \bar{\alpha}, \quad i = 1,2.
\end{split}
\end{equation}

Although in the unforced system (\( c_i = 0 \)) we have shown that \( \dot{V}(\boldsymbol{\zeta}) \leq 0 \) (see~\eqref{eq:d_v_3}), it is not possible to establish the existence of a positive definite continuous function \( P(\boldsymbol{\zeta}) \) such that \( \dot{V}(\boldsymbol{\zeta}) < -P(\boldsymbol{\zeta}) \). Consequently, \( V(\boldsymbol{\zeta}) \) does not qualify as a common Lyapunov function across all switching modes~\cite[Section~2.1]{liberzon2003switching}\cite{liberzon2004common}. Nevertheless, in each individual mode of the switching system, one state remains constant while the other decreases, ensuring that \( V(\boldsymbol{\zeta}) \) is non-increasing even under switching  (see~\cite[Section~3.1]{liberzon2003switching}), this is sufficient to guarantee internal stability, though asymptotic stability cannot be concluded. In the following, we derive an explicit upper bound on the allocation states and directly prove internal stability.

The general solution to the switching system~\eqref{eq:zeta1_dynamics} can be expressed as
\begin{equation} \label{eq:zeta1_solution_split}
    \zeta_i(t) = 
    \underbrace{\zeta_i(0) e^{ - \int_0^t b_{i\sigma(\xi)} \, d\xi }}_{\beta_{i1}}
    + 
    \underbrace{c_i \int_0^t e^{ - \int_s^t b_{i\sigma(\xi)} \, d\xi } b_{i\sigma(s)} \bar{\tau}^N(s) \, ds}_{\beta_{i2}}.
\end{equation}


\noindent Assuming the nominal control input \(\bar{\boldsymbol{\tau}}(t)\) is uniformly bounded for all \(t \geq 0\), i.e., there exists a constant \(\bar{\tau}_{\max} \succ 0\) such that \(\|\bar{\boldsymbol{\tau}}(t)\| \leq \bar{\tau}_{\max}\), it follows that \(|\tau^N(t)| \leq \sqrt{3}\bar{\tau}_{\max}=\bar{\tau}^N_{\max}\). 
Then, the second term \(\beta_{i2}\) can be bounded as
\begin{equation} \label{eq:beta2_bound_1}
    |\beta_{i2}| \leq |c_i| \bar{\tau}^N_{\max} \left| \int_0^t e^{ - \int_s^t b_{i\sigma(\xi)} \, d\xi } b_{i\sigma(s)} \, ds \right|,
\end{equation}
\noindent the integral inside the absolute value simplifies as
\begin{equation} \label{eq:kernel_integral}
    \int_0^t e^{ - \int_s^t b_{i\sigma(\xi)} \, d\xi } b_{i\sigma(s)} \, ds = 1 - e^{ - \int_0^t b_{i\sigma(\xi)} \, d\xi }.
\end{equation}
\noindent Since \(b_{i\sigma} \in \{0,\, k_i(w_i(t) + w_3(t)) > 0\}\), it follows that
\begin{equation} \label{eq:kernel_integral_bounds}
    (1 - e^{ - \int_0^t b_{i\sigma(\xi)} \, d\xi }) \in [0, 1),
\end{equation}
and thus,
\begin{equation} \label{eq:beta2_bound_final}
    |\beta_{i2}| \leq |c_i| \bar{\tau}^N_{\max} \left(1 - e^{ - \int_0^t b_{i\sigma(\xi)} \, d\xi } \right),
\end{equation}
\noindent similarly, the first term satisfies
\begin{equation} \label{eq:beta1_bound}
    |\beta_{i1}| \leq |\zeta_i(0)| e^{ - \int_0^t b_{i\sigma(\xi)} \, d\xi }.
\end{equation}

Combining these bounds gives
\begin{equation} \label{eq:zeta1_combined_bound}
    |\zeta_i(t)| \leq |\zeta_i(0)| e^{ - \int_0^t b_{i\sigma(\xi)} \, d\xi } + |c_i| \bar{\tau}^N_{\max} \left(1 - e^{ - \int_0^t b_{i\sigma(\xi)} \, d\xi } \right).
\end{equation}
Moreover, since \(b_{i\sigma} \in \{0,\, k_i(w_i(t) + w_3(t)) > 0\}\), it is straightforward to show that the following uniform bound holds
\begin{equation} \label{eq:zeta1_uniform_bound}
    |\zeta_i(t)| \leq \max \left\{ |\zeta_i(0)|,\, |c_i| \bar{\tau}^N_{\max} \right\},
\end{equation}
\noindent this guarantees that all signals within the allocation dynamics remain bounded for any bounded external input \(\boldsymbol{\tau}^N\), ensuring the internal stability of the switching allocation dynamics.

\addtolength{\textheight}{-0cm}   

\bibliographystyle{IEEEtran}
\bibliography{include/References}




%




\end{document}